\documentclass{aa}

\usepackage{graphicx}
\usepackage{amsmath}
\usepackage{txfonts}
\usepackage{booktabs}
\usepackage{xcolor}
\usepackage{hyperref}
\usepackage{orcidlink}
\AtBeginDocument{\nolinenumbers\let\linenumbers\relax}   
\definecolor{linkblue}{rgb}{0.08, 0.20, 0.50}
\hypersetup{colorlinks=true, linkcolor=linkblue, citecolor=linkblue, urlcolor=linkblue}

\newcommand{\Gaia}{\textit{Gaia}}
\newcommand{\getgaia}{\textsc{GetGaia}}
\newcommand{\masyr}{\,{\rm mas\,yr^{-1}}}
\newcommand{\mas}{\,{\rm mas}}
\newcommand{\kpc}{\,{\rm kpc}}
\newcommand{\kms}{\,{\rm km\,s^{-1}}}
\newcommand{\pmra}{\mu_{\alpha*}}
\newcommand{\pmdec}{\mu_{\delta}}
\newcommand{\parallax}{\varpi}

\providecommand{\degree}{^{\circ}}

\newcounter{apxnext}   

\begin{document}

\title{\getgaia{} 2.0: automatic membership and systemic motions of stellar systems with \Gaia{}}
\titlerunning{\getgaia{} 2.0: automatic membership and systemic motions with \Gaia{}}

\author{Andr\'es del Pino\orcidlink{0000-0003-4922-5131}\inst{1}\corrauth{apino@iaa.csic.es}}
\authorrunning{A. del Pino}

\institute{Instituto de Astrof\'isica de Andaluc\'ia (IAA-CSIC), Glorieta de la Astronom\'ia s/n, E-18008 Granada, Spain}

\date{Received \today; accepted ...}

\abstract
{\Gaia{} provides the astrometry needed to measure the systemic proper motions (PMs) of the stellar systems of the Local Group. Separating their member stars from the Milky Way foreground, however, still demands dedicated work for each object.}
{We present \getgaia{} 2.0, a public Python package that retrieves the \Gaia{} data around a stellar system and selects its member stars. The selection uses all the astrometric information of every source, with its full covariance matrix, together with the broad-band photometry and, when available, the line-of-sight velocities. \getgaia{} 2.0 is designed to run automatically, but it can be extensively configured by users for each specific scientific use.}
{The membership is a probabilistic model fitted by expectation maximization. It combines the astrometry, the color-magnitude diagram and the sky distribution of the stars, each with its own uncertainties and correlations, and the line-of-sight velocities of the stars that have one, from \Gaia{} or from an external catalog. The local systematic errors of the \Gaia{} reference frame are measured on the quasars of each field. The code returns the systemic PM and parallax with their full covariance, a detection significance, a contamination estimate, and the structural parameters of the system. We validate it on synthetic catalogs and run it on \Gaia{} DR3 for the known dwarf galaxies of the Local Group within 850\,kpc and for 95 Galactic globular clusters, in tiers that add one piece of literature information at a time, from the position alone to a dedicated run.}
{On the synthetic catalogs the code recovers the members with 90--100 percent purity. On \Gaia{} DR3,
it isolates the members and measures the systemic PMs of 55 dwarf galaxies and 93 globular clusters at a significance of at least $3\sigma$ over the field; 29 of the dwarfs and 86 of the clusters are recovered from their position alone. The dwarfs include every Milky Way satellite within 300\,kpc brighter than $M_V = -5.4$, ultra-faint systems down to $M_V = -0.8$, and nine galaxies at 370--820\,kpc. The PMs agree with the published values within $0.1\masyr$ for 47 of the dwarfs and within $0.1\masyr$ for all the clusters (median differences of $0.032$ and $0.009\masyr$), with uncertainties equal to those of dedicated studies, and no system deviates by more than $3\sigma$. The \Gaia{} line-of-sight velocities alone give the systemic velocity of 71 clusters, within $1\kms$ of the spectroscopic compilations in the median.}
{\getgaia{} 2.0 delivers, without manual intervention, the systemic motions of the stellar systems within the reach of \Gaia{}, with a complete and correlated error budget, together with their individual members, on which studies of their internal kinematics can build. All release-dependent quantities are isolated in a registry, so that the code is ready for the forthcoming \Gaia{} data releases.}

\keywords{astrometry -- proper motions -- galaxies: dwarf -- globular clusters: general -- Local Group -- methods: data analysis}

\maketitle

\section{Introduction}\label{sec:intro}

The \Gaia{} mission \citep{GaiaMission2016} has transformed the study of the stellar systems around the Milky Way (MW). Its successive data releases, the latest being DR3 \citep{GaiaDR3}, provide positions, parallaxes and proper motions (PMs) for more than a billion sources down to $G \sim 21$. They have allowed the measurement of the systemic motions of essentially all known MW globular clusters \citep[GCs; e.g.][]{Helmi2018, Vasiliev2019, Vasiliev2021}, dwarf galaxies \citep[e.g.][]{Helmi2018, Fritz2018, Simon2018, Kallivayalil2018, MassariHelmi2018, McConnachie2020a, McConnachie2020b, Li2021, Battaglia2022, Pace2022} and open clusters \citep{CantatGaudin2018, HuntReffert2023}, the detection of internal motions in the nearest of them \citep{Martinez-Garcia2021, delPino2021}, and the discovery of new satellites and streams \citep{Koposov2017, Torrealba2019, Malhan2018, PriceWhelan2018, Ibata2021}. Combined with the astrometry of the \emph{Hubble Space Telescope} (HST), which extends the time baseline and reaches fainter stars, \Gaia{} has gone beyond its own capabilities: the systemic PMs of distant dwarfs and the internal motions of the nearest ones are now within reach \citep{Massari2018, Massari2020, delPino2022, Warfield2023, Bennet2024, McKinnon2024}.

All these results rest on the same first step: separating the stars of the system from the overwhelming foreground of MW stars. A member star of a distant satellite is faint, its parallax is compatible with zero, and its PM is close to the bulk PM of the system, but so are the PMs of many halo stars. The selection is therefore a statistical problem in which the astrometric, photometric and spatial information must be combined. The heteroscedastic nature of the \Gaia{} errors cannot be ignored either. Bright stars at the tip of the red giant branch (RGB) of a satellite galaxy are precisely measured at $G \sim 16$, while for the horizontal branch (HB) stars at $G = 20$ the PM uncertainty of a single star is of the order of $0.5\masyr$, comparable to the separation between the bulk motion of the galaxy and the field stars in PM space. Several approaches have been proposed, from iterative sigma clipping in PM--parallax space \citep{Helmi2018, Fritz2018} to Bayesian mixture models combining astrometry, photometry and position \citep{Pace2019, McConnachie2020a, Battaglia2022, Pace2022, Jensen2024}, as well as methods tailored to GCs \citep{Vasiliev2021}.

\getgaia{}\footnote{\url{https://github.com/AndresdPM/GetGaia}} was written in this context as a simple, automated tool to download the \Gaia{} data around a given object and to produce a clean list of members. It was used in \citet{delPino2021} and \citet{Martinez-Garcia2021} to study the internal kinematics of the Sagittarius and the classical dwarf spheroidal (dSph) galaxies. That version relied on sequential cuts (a likelihood ratio in the color-magnitude diagram, a Gaussian fit with sigma clipping in PM--parallax space and, optionally, a multi-Gaussian expansion of the spatial distribution in the sky) and on several external packages that are no longer maintained.

With \Gaia{} DR4 scheduled for December 2026, we have rewritten the code from scratch. \getgaia{} 2.0 keeps the automated, name-to-catalog philosophy of the original, but replaces the sequential cuts by a single probabilistic model fitted by expectation maximization (EM). Every star enters the model with its own covariance matrix, built from the \Gaia{} uncertainties and correlation coefficients, and with its own photometric errors and, when it has one, the error of its line-of-sight velocity ($v_{\rm los}$). A faint star with a large error is thus neither rejected for being far from the systemic motion nor accepted for being compatible with everything: its weight is the one dictated by its likelihood ratio. Finally, the systematic errors of the \Gaia{} reference frame are measured on the quasars of each field and propagated to the systemic motion with their full covariance.

This paper is organized as follows. Section~\ref{sec:method} describes the data retrieval and the membership model. Section~\ref{sec:validation} validates the code on synthetic data. Section~\ref{sec:application} illustrates the selection on four systems of different nature (the Fornax dSph, the GC NGC\,5466, the ultra-faint Hercules and the open cluster M67). It then applies the code to the dwarf galaxies of the Local Group and to the MW globular clusters in tiers that add one piece of literature information at a time, from the position alone to a dedicated run, and compares the results with the published PMs and uncertainties. Section~\ref{sec:conclusions} summarizes the work. The appendices contain the tables of systemic motions, additional figures, the structural parameters returned by the fit, and the installation and basic use of the code.

\section{Method}\label{sec:method}

\begin{figure*}
\begin{center}
\includegraphics[width=\textwidth]{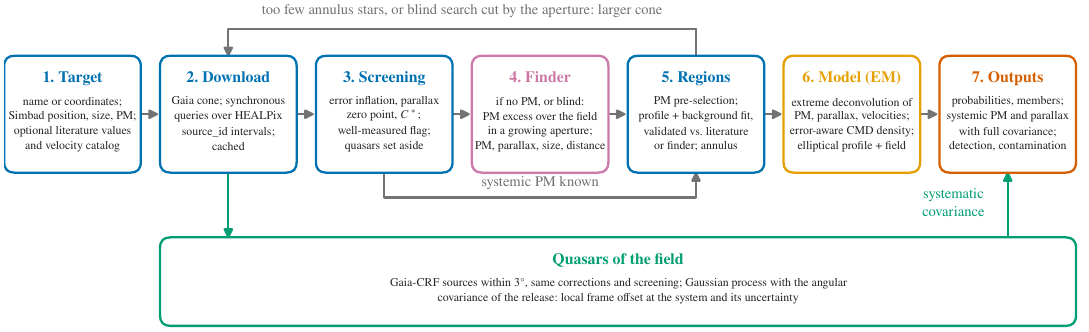}
\caption{Flow of a \getgaia{} 2.0 run. The main track goes from the target resolution (blue) to the membership model (orange) and the outputs (red); the side track (green) measures the local offset of the \Gaia{} reference frame on the quasars within three degrees of the target and feeds it, with its uncertainty, to the error budget of the systemic motion. The loop above the main track enlarges the cone when the control annulus holds too few stars or the blind search is cut by the aperture. The finder step (purple) runs when no systemic PM is known or in blind mode. The steps are those of Sects.~\ref{sec:retrieval} to \ref{sec:members}.}
\label{fig:flowchart}
\end{center}
\end{figure*}

Figure~\ref{fig:flowchart} summarizes the pipeline. \getgaia{} 2.0 is a pure Python package built on \textsc{numpy} \citep{Harris2020}, \textsc{scipy} \citep{Virtanen2020}, \textsc{pandas} \citep{McKinney2010}, \textsc{matplotlib} \citep{Hunter2007}, \textsc{astropy} \citep{astropy2022} and \textsc{astroquery} \citep{Ginsburg2019}. A run consists of three stages. The first stage retrieves the data (Sect.~\ref{sec:retrieval}). It resolves the target and downloads the cone, applies the calibration corrections and the quality screening, and sets aside the background quasars. The second stage finds the system (Sect.~\ref{sec:detection}). When no systemic PM is known, or when the user asks for it, the code locates the system in PM space from its position alone and estimates its distance. The third stage isolates the members (Sect.~\ref{sec:members}). It defines the object and field regions, fits the membership model, and derives the systemic motion with its error budget. Each step is a function that can also be called on its own from Python. \getgaia{} has two automatic modes. In the default mode, the code takes from Simbad everything it lists about the target and uses it as the starting point of the selection. In blind mode, it uses nothing but the name or the position of the target, and the finder of the second stage provides the starting point from the data themselves.

\subsection{Data retrieval}\label{sec:retrieval}

\subsubsection{Target resolution and download}\label{sec:data}

The target is identified by name. Its coordinates, systemic PM and angular size come from Simbad, as does the distance when Simbad lists measurements of it (the code takes their median), and the user can override any of them. When Simbad lists no PM, the code asks for one in an interactive run, takes the value given as an option, or finds the system by itself in the second stage of the run. A blind run keeps only the position and leaves everything else free. By default the \emph{search radius} is twice the angular diameter listed in Simbad, which encloses the bulk of a dSph galaxy or GC, and the downloaded cone is three times larger, so that a wide control field surrounds the object. A loose box in PM ($\pm 5\masyr$ around the systemic value, when known) and in parallax ($-5 < \parallax < 2\mas$) is applied on the server to reduce the volume of data. These cuts are deliberately wide to include members and to allow for a better modeling of the control field. The download stops at $G = 21$ by default (option \texttt{--max-gmag}). \Gaia{} goes somewhat deeper, but the astrometry of fainter sources carries little membership information, and we adopt this limit throughout the paper.

The ESA \Gaia{} archive limits the size and duration of a query and queues the asynchronous jobs. \getgaia{} avoids them: a first synchronous query lists the HEALPix pixels of the cone and their source counts, and since the \Gaia{} identifiers encode the pixel, the cone is retrieved as a series of short synchronous queries over \texttt{source\_id} intervals of at most $2\times10^4$ sources, which keep the archive on its primary-key index. The intervals are downloaded concurrently and cached on disk, so that an interrupted download resumes, and no login is required (Appendix~\ref{apx:usage}).

\subsubsection{Calibration corrections and quality screening}\label{sec:quality}

All the quantities that depend on the data release live in what we call the release registry, a small table of constants and calibration functions with one entry per release: the table name and reference epoch, the systematic error floors of the astrometry, the inflation factors of the astrometric errors \citep[1.05 and 1.22 for five- and six-parameter solutions;][]{Fabricius2021}, the corrected flux excess factor $C^\ast$ and its expected scatter \citep{Riello2021}, the parallax zero point \citep{Lindegren2021b}, the quality thresholds, and the table and selection of quasars used to measure the local systematics. For DR3 all of them are implemented. For DR4 the registry holds the table name and epoch of the draft data model, and each calibration step is skipped with a warning until the DR4 validation papers fill in its values. Every correction can be disabled individually, and users can register their own release.

After the corrections, a source is flagged as well measured if ${\rm RUWE} \leq 1.4$, ${\tt ipd\_gof\_harmonic\_amplitude} \leq 0.2$, ${\tt visibility\_periods\_used} \geq 9$, ${\tt astrometric\_excess\_noise\_sig} \leq 2$ \citep{Lindegren2021a, Fabricius2021}, and $|C^\ast|$ lies within three times its expected scatter at its magnitude \citep{Riello2021}. All the thresholds are configurable. The screening removes mostly faint or blended sources, a sizable fraction of the stars above the membership threshold in the crowded centers of nearby systems.

\subsubsection{Background quasars and galaxies}\label{sec:quasars}

Quasars and compact galaxies have the astrometric signature of a distant stellar system, a parallax and PM compatible with zero, and at high Galactic latitude they can outnumber the members of a distant satellite. We do not remove them with the \Gaia{} classifications of extragalactic sources \citep[the DSC probabilities of][and the \texttt{qso\_candidates} table]{DelchambreDSC2023}, which rely in part on the astrometry and flag a sizable fraction of the genuine faint members as quasars; the membership model handles them instead, with a field component for distant sources. The one exception is the set of confirmed quasars of the \Gaia{} celestial reference frame \citep[\Gaia{}-CRF3, flagged as \texttt{gaia\_crf\_source} in \texttt{qso\_candidates};][]{Klioner2022}, a sample essentially free of stars. This sample is set aside from the fit and from the member list and used to measure the local systematic errors of the reference frame.

\subsection{Finding the system}\label{sec:detection}\label{sec:finder}\label{sec:distance}

The second step is to find the system in the data. By default, \getgaia{} takes from Simbad as much information as it lists, the systemic PM, the size and the distance, as the starting point of the selection; the user can override any of these values. When a value is missing, or in blind mode, \getgaia{} derives it from the data, by contrasting the stars around the target with those of the surrounding field in PM space. We call this stage the finder. It isolates no star: it works on the excess of the inner region over the field and returns a first measurement of the systemic PM, parallax, size and distance, which the membership model then refines. The finder works as follows.

The stars within an aperture around the target are mapped in the PM plane, in cells of $0.25\masyr$, each as a Gaussian kernel whose width is the quadrature sum of its PM error and a floor of $0.35\masyr$. The same map, built for the stars beyond the search radius and scaled by the ratio of the areas, is subtracted, and the difference divided by its Poisson error gives the significance of the excess in each cell. The aperture grows from a tenth of the search radius to the whole of it in eight steps, so that a small system stands out inside its own radius; in blind mode the search radius itself doubles, from $0.5\degree$ to $1\degree$, while no excess is found or while the excess extends to the aperture. The significance is a ranking statistic, not a detection: the detection test of the third stage decides afterwards whether there is a system at all. Every local maximum above a threshold ($3\sigma$ by default) is a candidate, and up to five are kept, ranked by significance.

For each candidate, the systemic PM is an iterative weighted mean of the stars within $0.75\masyr$ (plus their errors) of the current estimate, weighted by the fraction of the excess in their cell and by their inverse squared errors; its error is that of the weighted mean or the scatter of the stars, whichever is larger. The parallax is the median of the same stars, and the radius that holds half of the excess counts is the size. The finder also estimates the distance of the candidate, from the parallax when it exceeds $0.05\mas$ at $5\sigma$ and otherwise from the tip of the RGB: the luminosity function of the excess is scanned from bright to faint, and the magnitude at which the cumulative excess reaches five stars at $3\sigma$, followed by at least 20 stars within 1.5\,mag, is taken as the tip, at $M_G = -3.5$. Compared with the literature distances of the systems that we recover from the position alone in this paper, the parallax estimate is accurate (median ratio 1.03, 16th to 84th percentiles 0.84--1.20). The accuracy of the RGB onset depends on how well populated the upper RGB is: the ratio has a median of 1.2 for the dwarf galaxies but of 3 for the globular clusters, whose few stars near the tip put the onset well below it, and it can be off by an order of magnitude for the sparsest systems. This is enough for its only use, a bound on the intrinsic velocity dispersion of the system in the membership model, which an overestimated distance only tightens. No distance is returned when both methods fail. These finder-level values, PM with its error, parallax, significance, size and distance, are the first of the two measurements that \getgaia{} reports.

A candidate is, by construction, a group of co-moving stars at the distance that the finder estimates for it, whether it is the target or a foreground group, so that distance cannot tell the two apart. The distance of the target is what separates them, and it must come from outside, from Simbad or from the user. When the distance is known, the candidates whose PM would imply a tangential velocity above $1000\kms$ at that distance (a limit the user can change), or whose parallax is incompatible with it, are discarded. The same bound is applied again to the systemic PM of the final membership, which can drift from the candidate toward a richer population, and a membership that fails it is not reported as a detection. If several candidates remain, the code asks the user and takes the strongest by default; all of them are kept in the output. Without a distance, the strongest candidate is taken, and the membership model and its detection test are the only safeguard. When no cell reaches the threshold, the run proceeds without a guess. In blind mode the membership model still runs, so that the diagnostic outputs are produced, but its result is not reported as a detection, since without a guess the model follows whatever population dominates the cone.

\subsection{Isolating members stars}\label{sec:members}

The third and last step isolates the member stars with a probabilistic model fitted by EM, starting from the prior information of Simbad or the user (Sect.~\ref{sec:data}) or, when there is none or in blind mode, from the finder (Sect.~\ref{sec:detection}).

\subsubsection{Object and field regions, and structural parameters}\label{sec:regions}

\getgaia{} contrasts the stars of the region that contains the object with those of a control field around it, far enough not to contain its stars. The field model is learned from this control field, so the boundary between the two must follow the size and shape of the system, and it must be drawn before any membership probability exists. The code proceeds in three steps.

\paragraph{Step 1: the preliminary profile.} An elliptical exponential (or Plummer) profile plus a uniform background is fitted by EM to the positions of the stars whose PM lies within $1\masyr$ (plus three times their error) of the initial guess, the Simbad or user value, or the PM of the finder. This fit is relatively fragile: in fields with a strong density gradient or a nearby system, the profile can grow to its size bound or wander away from the target. The code therefore validates it and, if needed, replaces it. Literature values of the half-light radius, ellipticity and position angle, when available, provide the starting point and the reference: a fit is rejected if the profile reaches its size bound, if its half-light ellipse does not contain the target, if its center falls more than one half-light radius from the target, or if its size differs from the literature value by more than a factor of three. A rejected fit is repeated with the center fixed on the target; if this also fails, the literature shape is used as it is. In blind mode, or when there are no literature values and the finder ran, its half-excess radius, as a circle, plays the role of the literature shape throughout, with a size tolerance of a factor of two instead of three, since it is only a rough size; when neither is available, the fallback is a circle of $0.15\degree$.

\paragraph{Step 2: the regions.} The object region is the ellipse within which the preliminary profile exceeds a given fraction of the background density (10\% by default), capped at five half-light radii so that the object region of a compact system does not fill with field stars, and the control annulus is the rest of the cone; both the fraction and the boundary can be set by the user. Two safeguards enlarge the cone and download it again, up to a radius of $3\degree$. If the control annulus holds fewer than 500 well-measured stars, too few to constrain the field model, the cone grows, as happens for compact systems at high Galactic latitude. If the object boundary exceeds half of the cone radius, the cone grows so that an extended system does not leave only a thin ring for the control annulus polluted by member stars. The regions are defined even when the spatial term is not used in the final probabilities, because the control annulus is needed to model the field.

\paragraph{Step 3: the final profile and the second pass.} The membership model is fitted with the preliminary profile as the starting point of its spatial term. The profile is refitted at every EM step with the membership probabilities as weights, whenever at least 30 members are expected, and when the EM converges the parameters of the last profile (half-light radius, ellipticity and position angle) are reported as the structural parameters of the system. This step gives a second chance to the systems whose preliminary fit was replaced in step 1, since the probability-weighted sample is much cleaner than the pre-selected one: the free fit is tried again and adopted if it passes the same validation and keeps its half-light radius within a factor of 1.5 of the literature one, or of the finder's size in its place. The output records how the final profile was obtained. Finally, the boundary of step 2 is recomputed if the final profile differs much from the preliminary one. Specifically, if more than 2 percent of the light of the final profile falls outside the boundary, in the control annulus, and the semi-major axis of the boundary that the same rule draws from the final profile differs from the original one by more than 10 percent, the regions are redrawn with the new boundary, which takes the shape of the final profile, and the model is fitted again, so that the field is always learned from an annulus free of members. When the user has fixed the boundary, the redrawn one is never smaller than that value.

\subsubsection{The membership model}\label{sec:model}

Let $\boldsymbol{a}_i = (\pmra, \pmdec, \parallax)_i$ be the astrometric vector of star $i$, with covariance matrix $\mathbf{S}_i$ built from the listed uncertainties and the three correlation coefficients provided by \Gaia{}, plus optional systematic floors added in quadrature to the diagonal; let $\boldsymbol{c}_i = (G_{\rm BP} - G_{\rm RP}, G)_i$ be its position in the color-magnitude diagram (CMD), with errors $\boldsymbol{\sigma}_i$; and let $\boldsymbol{r}_i$ be its position on the tangent plane of the sky. The stars in the cone are drawn from one of two populations, the object and the field, each described by the product of a spatial term, an astrometric term and a photometric term,

\begin{equation}\label{eq:likelihoods}
\begin{aligned}
\mathcal{L}_{{\rm obj},i} &= N_{\rm obj}\, f(\boldsymbol{r}_i)\, \mathcal{N}(\boldsymbol{a}_i\,|\,\boldsymbol{\mu}, \mathbf{V} + \mathbf{S}_i)\, K_{\rm obj}(\boldsymbol{c}_i), \\
\mathcal{L}_{{\rm field},i} &= \frac{N_{\rm field}}{A} \sum_{k=1}^{K} \alpha_k\, \mathcal{N}(\boldsymbol{a}_i\,|\,\boldsymbol{\mu}_k, \mathbf{V}_k + \mathbf{S}_i)\, K_{\rm field}(\boldsymbol{c}_i),
\end{aligned}
\end{equation}

\noindent and the posterior membership probability of star $i$ is

\begin{equation}\label{eq:posterior}
P_i = \frac{\mathcal{L}_{{\rm obj},i}}{\mathcal{L}_{{\rm obj},i} + \mathcal{L}_{{\rm field},i}},
\end{equation}

\noindent where $N_{\rm obj} = \sum_i P_i$ and $N_{\rm field} = \sum_i (1 - P_i)$ are the expected numbers of stars of each population, $A$ is the area of the cone, and $\mathcal{N}(\boldsymbol{x}\,|\,\boldsymbol{m}, \mathbf{C})$ denotes the multivariate normal density. The three terms of Eq.~\ref{eq:likelihoods} are, in order, the spatial term $f(\boldsymbol{r}_i)$ (uniform, $1/A$, for the field), the astrometric term $\mathcal{N}(\boldsymbol{a}_i\,|\,\cdot)$ and the CMD term $K(\boldsymbol{c}_i)$. We describe them in turn, followed by the optional $v_{\rm los}$ term.

\paragraph{Astrometry.} The object is a single Gaussian with mean $\boldsymbol{\mu}$ (the systemic PM and parallax) and intrinsic covariance $\mathbf{V}$, and the field is a mixture of $K$ Gaussians ($K = 4$ by default) with amplitudes $\alpha_k$, means $\boldsymbol{\mu}_k$ and covariances $\mathbf{V}_k$, fitted within the wide PM and parallax box of the download (Sect.~\ref{sec:data}), so that the probabilities are self-consistent. Each star is compared with the model through $\mathbf{V} + \mathbf{S}_i$, and the parameters are updated with the extreme deconvolution equations of \citet{Bovy2011}. The intrinsic PM dispersion of a dwarf galaxy is small, about $0.02\masyr$ for $10\kms$ at $100\kpc$ ($1\masyr$ is $4.74\kms$ at 1\,kpc), but it is left free so that it absorbs the small-scale systematics that are not in $\mathbf{S}_i$. A free component has a danger: with few members it can broaden into a second field component, or drift onto an unrelated clump of field stars. Two safeguards act in that regime. The intrinsic velocity dispersion is bounded, $\sigma_v \leq \sigma_{v,{\rm max}}$ ($30\kms$ by default), which caps the eigenvalues of $\mathbf{V}$ at $\sigma_{v,{\rm max}}/(4.74\,D)$ when the distance $D$ (in kpc) is known, from the user or from the finder, and at $1\masyr$ otherwise; and a weak Gaussian prior on the systemic PM ($0.5\masyr$, centered on the initial guess) holds the mean in place until the system has more than a few tens of members. A lower limit on the dispersion of the field components ($0.2\masyr$) keeps them from collapsing onto the object, and a further field component with zero PM and parallax and a fixed dispersion of $0.03\masyr$ represents the quasars and compact galaxies (Sect.~\ref{sec:quasars}), a narrow clump that the object component could otherwise absorb.

\paragraph{Color-magnitude diagram.} $K_{\rm obj}$ and $K_{\rm field}$ are non-parametric densities built from the stars themselves, weighted by $P_i$ and $1 - P_i$, with a Gaussian kernel per star whose width is the quadrature sum of its photometric errors and a smoothing floor of 0.02\,mag. No isochrone or stellar population model enters the selection at any stage: the shape of the sequence is learned from the data, so the model needs no assumption on the age, metallicity or distance of the system, and handles composite populations, young stars and the HB alike. The likelihood of a star is the population density convolved with its own error kernel, so a faint star with a 0.2\,mag color error is compared with a broader density than a bright giant. Each star is removed from its own density (leave-one-out), so that isolated outliers cannot support themselves, and the densities are evaluated on a grid, with the stars grouped by kernel width.

\paragraph{Sky position.} $f(\boldsymbol{r})$ is an elliptical exponential or Plummer profile with free center, ellipticity, position angle and scale radius, normalized over the cone, and the field is uniform. The profile is fitted by weighted maximum likelihood with the current probabilities as weights, from the preliminary profile of Sect.~\ref{sec:regions}. When the spatial term is disabled, $f$ is uniform like the field: the position of a star then carries no information and members can be found anywhere in the cone, for instance in the low-surface-brightness structures around a system. The uniform field is adequate for the cones of a few degrees used here, but not where the field density varies across the cone, close to the Galactic plane, the bulge or the Magellanic Clouds; a gradient of the field density is a natural extension, planned for a future version.

\paragraph{Line-of-sight velocity.} A fourth, optional term uses the line-of-sight velocities of the stars that have one: the \Gaia{} RVS velocities, which reach $G \simeq 14$ in DR3, and those of an external spectroscopic catalog, matched to the \Gaia{} sources by \texttt{source\_id} or by position at the epoch of the catalog (\getgaia{} performs the matching automatically if an external catalog is provided). A star with $v_{{\rm los},i} = v_i$ and error $e_i$ contributes a factor $\mathcal{N}(v_i\,|\,v_{\rm los,sys}, \sigma_{\rm los}^2 + e_i^2)$ to $\mathcal{L}_{{\rm obj},i}$ and $\mathcal{N}(v_i\,|\,v_{\rm field}, \sigma_{\rm field}^2 + e_i^2)$ to $\mathcal{L}_{{\rm field},i}$; a star without one is evaluated exactly as before. The systemic velocity $v_{\rm los,sys}$ and the intrinsic dispersion $\sigma_{\rm los}$ are deconvolved from the errors, with $\sigma_{\rm los}$ capped at $\sigma_{v,{\rm max}}$. The field term is not learned from external catalogs, which usually target members and holds few, unrepresentative field stars, but from the Milky Way along the line of sight: a Gaussian at the reflex of the solar motion with a dispersion of $100\kms$, refined on the RVS stars of the control annulus when there are at least 30. A foreground interloper is thus rejected on its velocity alone, and the presence or absence of a $v_{\rm los}$ never counts as evidence. The term factorizes from the astrometry, so the covariance of the systemic motion is block diagonal, as expected for measurements from different instruments.

The EM scheme alternates the evaluation of Eqs.~\ref{eq:likelihoods} and \ref{eq:posterior} with the update of all the parameters, with a few extreme deconvolution iterations at every step, and stops when the systemic motion and the probabilities have stabilized, typically after 10--30 iterations. The model is fitted on the well-measured stars of Sect.~\ref{sec:quality} or, in the densest fields on a random subset of $\sim 10^5$ stars of the control annulus. The probabilities of the stars that fail the quality screening are then evaluated with the fitted model and reported separately, since in crowded systems many faint members fail the excess-noise and $C^\ast$ cuts. For every star, the output catalog contains the probability of each term alone, the total probability, the log-likelihood ratios, and a membership flag set by the chosen threshold.

\subsubsection{Systemic motion and error budget}\label{sec:errors}

The systemic PM and parallax are the mean $\boldsymbol{\mu}$ of the object component. Its statistical covariance is the inverse Fisher information of the mean,

\begin{equation}\label{eq:fisher}
\mathbf{C}_{\rm stat} = \left[\sum_i P_i\, (\mathbf{V} + \mathbf{S}_i)^{-1}\right]^{-1},
\end{equation}

\noindent which weights each star by its probability and by its full covariance, and naturally includes the intrinsic dispersion. As the inverse of the Fisher information, $\mathbf{C}_{\rm stat}$ inherits the correlations between the astrometric parameters of the individual \Gaia{} solutions. The total covariance of the systemic motion, statistical plus systematic, is reported in full (three by three, for the two PM components and the parallax), so that the correlation between $\pmra$ and $\pmdec$, which is often far from zero, can be propagated into orbit integrations.

For systems with thousands of members the statistical errors are far smaller than the systematic errors of the \Gaia{} reference frame, which are correlated on the sky: in DR3 the angular covariance of the quasar PMs has a large-scale term of $292\,(\mu{\rm as\,yr}^{-1})^2$ with an e-folding angle of $12\degree$ \citep[][Sect.~5.6]{Lindegren2021a}, and with the small-scale term of our covariance function its value at zero separation is $550\,(\mu{\rm as\,yr}^{-1})^2$, i.e. $23\,\mu{\rm as\,yr}^{-1}$ for a compact system, close to the $0.026\masyr$ floor of VB21. \getgaia{} measures the local part of this error on the \Gaia{}-CRF3 quasars (Sect.~\ref{sec:quasars}) within $3\degree$ of the target, about a thousand at their mean density of 38 per square degree, corrected and screened like the stars and clipped at $5\sigma$. Since quasars have no detectable motion, their astrometry samples the systematic field of the reference frame, $\boldsymbol{s}(\boldsymbol{r})$, at their positions. We model each component of $\boldsymbol{s}$ as a Gaussian random field with zero mean and the angular covariance $V(\theta)$ of the release, for DR3 the large-scale terms of \citet{Lindegren2021a} plus a small-scale term of the kind introduced by VB21 and adopted by \citet{Battaglia2022}, here an exponential with a $0.25\degree$ scale and the amplitude that brings $V(0)$ to $550\,(\mu{\rm as\,yr}^{-1})^2$ and $700\,\mu{\rm as}^2$:

\begin{align}\label{eq:kernel}
V_\mu(\theta) &= 292\,e^{-\theta/12\degree} + 258\,e^{-\theta/0.25\degree}\ (\mu{\rm as\,yr}^{-1})^2, \nonumber\\
V_\varpi(\theta) &= 142\,e^{-\theta/16\degree} + 558\,e^{-\theta/0.25\degree}\ \mu{\rm as}^2 .
\end{align}

\noindent A Gaussian-process regression then gives the posterior of the field averaged over the footprint of the members, with the weights of Eq.~\ref{eq:fisher}. The posterior mean $\boldsymbol{\delta}$ is the local offset of the reference frame at the system, measured to about $0.01\masyr$ with a thousand quasars, and the posterior covariance $\mathbf{C}_{\rm QSO}$ is the systematic uncertainty that the quasars leave. Without quasars near the system the posterior tends to the prior, the all-sky floor averaged over the footprint, which lies below $23\,\mu{\rm as\,yr}^{-1}$ for extended systems because the small-scale term averages out. If the quasar residuals exceed their errors (reduced $\chi^2$ above one), the errors are scaled up before the final solution, and with fewer than ten usable quasars the systematic term is taken from the default floors of the release registry, $0.034\masyr$ per PM component for DR3, above the zero-separation value of Eq.~\ref{eq:kernel}. The total covariance of the systemic motion is

\begin{equation}\label{eq:total}
\mathbf{C}_{\rm tot} = \mathbf{C}_{\rm stat} + \mathbf{C}_{\rm QSO} + \boldsymbol{\delta}\boldsymbol{\delta}^{\rm T},
\end{equation}

\noindent where the last term is included because the systemic motion is not corrected for the offset by default, so that the quoted error covers the expected squared error of the uncorrected value. When the user asks \getgaia{} to subtract the offset from the systemic motion (option \texttt{--quasar-correct}), that term is dropped.

When line-of-sight velocities are used, the statistical error of $v_{\rm los,sys}$ is the inverse Fisher information of the $v_{\rm los}$ term over the stars with a velocity, and the systematic term is the zero-point uncertainty of their source: $0.3\kms$ for the \Gaia{} RVS, the level of the zero-point differences between the RVS and the ground-based catalogs reported by \citet{Katz2023}, or the value supplied for an external catalog. The output reports the joint four-dimensional statistical and total covariance matrices of $(\pmra, \pmdec, \parallax, v_{\rm los})$.

Finally, the code reports a detection significance and a contamination estimate that do not rely on the probabilities being calibrated. The membership found within an aperture of 2.5 half-light radii of the profile (or the object region, if smaller), $\sum_i P_i$, is compared with the membership the model would assign to the stars of the control annulus if they were placed in that aperture, scaled by the area ratio. The excess in units of its Poisson error is the significance, and the number of annulus stars that would pass the membership threshold, scaled the same way, is the expected number of contaminants in the member list, estimated from the field itself. A significance below 3 means that the object component has settled on field stars rather than on a real population, and the members are reported as unreliable; when no star passes the threshold, the systemic motion is reported as undefined. The test assumes that the field population is the same inside and outside the object region, so moderate values (3--5) deserve caution in fields with strong gradients. As a last safeguard, \getgaia{} compares the systemic PM of the members with that of the finder (Sect.~\ref{sec:finder}); when the two differ by more than $3\sigma$ and $0.1\masyr$, the run warns that the membership may be contaminated.

\section{Validation with synthetic data}\label{sec:validation}

\begin{figure}
\begin{center}
\includegraphics[width=\columnwidth]{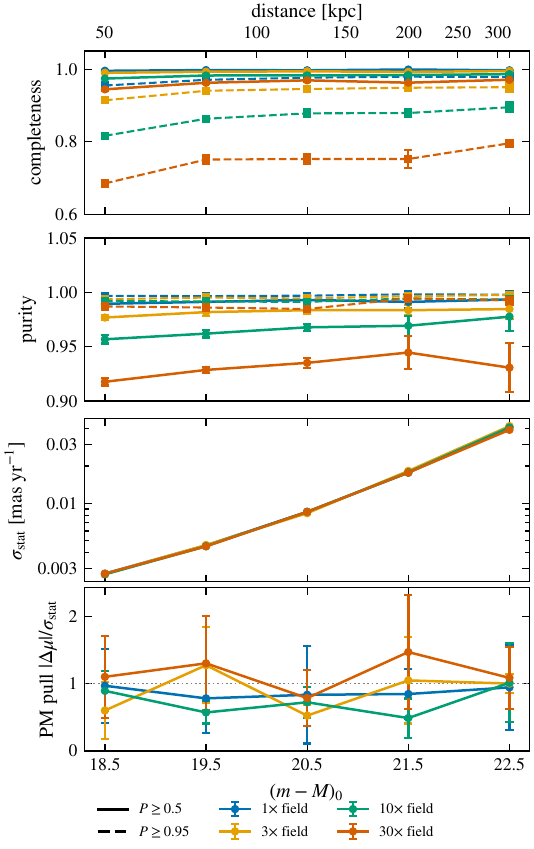}
\caption{Performance of the full model on the mock catalogs of Sect.~\ref{sec:validation} as a function of the distance modulus of the system, for field densities of 1, 3, 10 and 30 times that of the reference field (see text). From top to bottom: completeness of the selection at $P \geq 0.5$ (solid) and $P \geq 0.95$ (dashed); purity; statistical error of the recovered systemic PM (quadrature mean of the two components); and deviation of the recovered PM from the input in units of that error. Points and error bars are the mean and standard deviation over five realizations.}
\label{fig:study}
\end{center}
\end{figure}

\begin{figure}
\begin{center}
\includegraphics[width=\columnwidth]{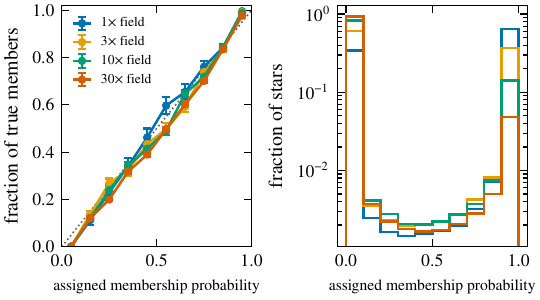}
\caption{Calibration of the membership probabilities over the 100 mock catalogs, per field density. Left: fraction of true members among the stars assigned a given probability (binomial error bars); a calibrated model follows the dotted diagonal. Right: distribution of the assigned probabilities.}
\label{fig:calibration}
\end{center}
\end{figure}

We have validated \getgaia{} on synthetic catalogs. The package includes a mock generator (\texttt{getgaia.simulate}). It draws the stars of a system from a PARSEC isochrone \citep{Bressan2012} in the \Gaia{} passbands following the initial mass function of \citet{Kroupa2001}, places them at a given distance modulus, and assigns them an elliptical exponential profile and a small intrinsic velocity dispersion. The astrometry and photometry are then perturbed with the \Gaia{} DR3 uncertainties as a function of magnitude from the \textsc{PyGaia} package, including random correlation coefficients. The field can be synthetic or, more realistically, resampled from the stars of a real \Gaia{} field.

We built a systematic set of mock catalogs that reproduce the actual conditions of a \Gaia{} membership analysis. The system is a 10\,Gyr, ${\rm [Fe/H]} = -1.5$ population with a stellar mass of $2.3 \times 10^6\,{\rm M}_\odot$ (4000 stars brighter than $G = 21$ at 80\,kpc), a half-light radius of 250\,pc, an ellipticity of 0.3, a velocity dispersion of $10\kms$ and a systemic PM of $(+0.38, -0.36)\masyr$. It is placed at distance moduli from 18.5 to 22.5 (50 to 316\,kpc). The tip of the RGB then moves from $G = 15.4$ to $G = 19.4$, the number of members brighter than $G = 21$ drops from 6600 to 270, and their PM errors grow from $0.03$ to $0.7\masyr$. The field is resampled, with replacement and a small random perturbation of the astrometry, from the 7534 \Gaia{} stars of the control annulus of a cone around the Fornax dSph, a high-latitude field ($b = -65\degree$) that holds 413 stars per square degree inside the PM box, so that it has the true distribution of astrometry, photometry and errors of a halo field at any density. The mock cones, of $1\degree$ radius or five half-light radii if larger, are filled at 1, 3, 10 and 30 times the density of the real field; the last value is comparable to a field at low Galactic latitude. The grid therefore spans five distances and four densities, and the contamination inside the half-light radius grows with the density from 1--3\% at the base value to 24--45\% at thirty times it. Five realizations are run per cell of the grid, with the full model and a PM guess offset by $0.1\masyr$ from the truth, and the results are averaged. Since the CMD term is non-parametric, the age and metallicity of the mock population are of no consequence for the model: a young or composite population is treated exactly as the old population simulated here.

Table~\ref{tab:grid} and Fig.~\ref{fig:study} summarize the 100 runs. At $P \geq 0.5$ the completeness stays above 94\% and the purity above 92\% in every cell, the lowest values at thirty times the base density. Raising the threshold to $P \geq 0.9$ keeps the purity at 98--100\% everywhere for a completeness of 78--99\%, lowest in the densest fields, and $P \geq 0.95$ buys a further half percent of purity at a larger cost in completeness. The systemic PMs are recovered with a statistical error that grows from $0.003\masyr$ at 50\,kpc to $0.04\masyr$ at 316\,kpc, as the members drop from 6600 to 270, and without bias: over the 100 runs the deviation from the input, in units of that error, has a root-mean-square of 1.05 and a mean consistent with zero in both components. The half-light radius is recovered within 5\% in all cells. The behavior is remarkably flat with distance. The sparse systems at 300\,kpc, with only 270 members whose PM errors are comparable to the separation from the field, are recovered as well as the nearby ones, because the astrometric term weights each star by its own covariance and the photometric and spatial terms take over where the astrometry is uninformative.

\begin{table}
\caption{Recovery of the mock system over the grid of Sect.~\ref{sec:validation}, with the \Gaia{} DR3 and DR4 error models: for each field density (in units of that of the Fornax field), the range over the five distances of the contamination inside the half-light radius, of the completeness $C$ and purity $\Pi$ of the members selected at $P \geq 0.5$, $0.9$ and $0.95$, and of the deviation of the recovered systemic PM from the input in units of its statistical error, all in percent except the last and averaged over the five realizations of each cell. The statistical error itself depends on the distance and not on the density: from $0.003\masyr$ at 50\,kpc to $0.04\masyr$ at 316\,kpc with DR3, and from $0.0014$ to $0.02\masyr$ with DR4.}
\label{tab:grid}
\centering
\scriptsize
\setlength{\tabcolsep}{2.5pt}
\begin{tabular}{l r r r r r r r r r}
\toprule
 & Field & Cont. & $C_{0.5}$ & $\Pi_{0.5}$ & $C_{0.9}$ & $\Pi_{0.9}$ & $C_{0.95}$ & $\Pi_{0.95}$ & $|\Delta\mu|/\sigma$ \\
\midrule
DR3 & 1 & 1--3 & 100 & 99 & 97--99 & 100 & 95--98 & 100 & 0.8--1.0 \\
DR3 & 3 & 3--8 & 99 & 98 & 95--97 & 99 & 91--95 & 99--100 & 0.5--1.3 \\
DR3 & 10 & 9--22 & 97--99 & 96--98 & 88--94 & 99 & 82--90 & 99 & 0.5--1.0 \\
DR3 & 30 & 24--45 & 94--97 & 92--94 & 78--88 & 98--99 & 68--80 & 98--99 & 0.8--1.5 \\
\midrule
DR4 & 1 & 1--3 & 100 & 100 & 99--100 & 100 & 99--100 & 100 & 0.6--1.2 \\
DR4 & 3 & 3--8 & 100 & 99 & 99--100 & 100 & 97--99 & 100 & 0.5--1.3 \\
DR4 & 10 & 9--22 & 99--100 & 98--99 & 97--98 & 99--100 & 94--97 & 100 & 0.6--1.0 \\
DR4 & 30 & 24--45 & 98--99 & 97--98 & 93--97 & 99 & 88--94 & 99--100 & 0.7--1.1 \\
\bottomrule

\end{tabular}
\end{table}

The same grid run with the uncertainties expected for \Gaia{} DR4 (Table~\ref{tab:grid}; Fig.~\ref{fig:study_dr4}, Appendix~\ref{apx:dr4}) shows what the longer baseline will bring. With PM errors 2.3 times smaller at a given magnitude, the purity at $P \geq 0.5$ rises to 97--100\% in every cell, and the completeness at $P \geq 0.9$ in the densest fields from 78--88\% to 93--97\%. The statistical errors of the systemic PMs shrink by a factor of about two, to $0.0014$--$0.02\masyr$, and the deviations from the input keep a root-mean-square of $1.0\sigma$. The contaminated cases that are marginal today, a dwarf galaxy at 50\,kpc behind a disk-like field, become clean with DR4.

The dense fields also show the value of the quasar-like field component and of the bound on the scale radius of the spatial profile (Sect.~\ref{sec:model}) when the extragalactic sources outnumber the members: together they keep the object component on the system, with an unbiased systemic PM and a purity above 92\%, even for 280 members among 39\,000 field stars. Without them, the object component would absorb the clump of extragalactic sources, with a PM bias of up to $0.1\masyr$ and a purity below 60\%. Figure~\ref{fig:calibration} shows the calibration of the probabilities over all the runs. The probabilities are strongly bimodal: about 2\% of the stars have $0.1 < P < 0.9$. For those stars the probabilities are slightly optimistic, i.e. a star assigned $P = 0.5$ is a true member in about 45\% of the cases, so that $P$ should be read as a ranking rather than as an exact frequency in that range. At $P \geq 0.9$ the calibration is accurate, with 98--100\% true members. The purity at $P \geq 0.5$ is far above 50\% because the selection is dominated by the stars at $P \simeq 1$; the stars with intermediate probabilities are few, and they are the ones that a higher threshold removes. The extreme-deconvolution offset and scatter of a synthetic quasar sample with 2.5\% of interlopers are recovered after the clipping of Sect.~\ref{sec:errors}. A simpler test, on a bright sequence immersed in a Gaussian field, which also shows what each term of the model contributes, is given in Appendix~\ref{apx:toy}. All these tests, as well as unit tests of every module, run offline and are executed automatically on every change of the code.

The grid varies the density of the field but not its distribution in the PM plane, in which the system lies relatively away from the bulk of the field stars. The separation between the two depends on the line of sight, through the reflex of the solar motion, so we also ran the grid with the system placed on the PM peak of the field, both in this field and in that of Draco, at $b = +35\degree$ and ten times denser. There the astrometry alone cannot separate the members, and the cost falls on the completeness rather than on the purity: at $P \geq 0.95$ the purity stays at 97--100\% in the original field and at 90--99\% in the Draco field, against 98--100\% above, while the completeness drops to 57--99\% and 50--92\%, from 68--98\%, and the stars of intermediate probability become numerous (the purity at $P \geq 0.5$ falls to 84\% and 50\% in the densest cells). At 316\,kpc the system is no longer recovered beyond a few times the base density of the Draco field: the object component absorbs the field, and the contamination estimate of Sect.~\ref{sec:errors} flags the result. The systemic PM is pulled toward the field by up to $0.01$ and $0.04\masyr$ in the densest cells, a shift of the size of the systematic term of Sect.~\ref{sec:errors}, to be kept in mind for a populous system whose statistical error is far smaller. The real systems that follow, in fields of every kind, test that regime.
\section{Application to dwarf galaxies and globular clusters}\label{sec:application}

In this section we apply \getgaia{} to real stellar systems and obtain their systemic properties and their member stars. In its default mode the code takes from Simbad whatever it lists about a system, the PM, the size and the distance, using them as the starting point to isolate its member stars and skipping the finder. We wanted to test \getgaia{} fully, from the position alone and without any prior, and at the same time to obtain measurements that do not depend on previous works. We therefore ran it on every system in a sequence of four tiers, each adding one piece of literature information, and we keep for each system the result of the first tier that recovers it:

\begin{list}{}{\setlength{\labelwidth}{3.0em}\setlength{\leftmargin}{3.4em}\setlength{\labelsep}{0.4em}\setlength{\itemsep}{1pt}\setlength{\parsep}{0pt}\setlength{\topsep}{3pt}}
\item[Tier 1.] The position alone (blind mode, Sect.~\ref{sec:detection}): the finder provides the PM, the parallax, the size and the distance, and the membership model refines them.
\item[Tier 2.] The position and the literature distance, which bounds the intrinsic dispersion of the system and rejects the candidates of the finder that would imply an impossible velocity.
\item[Tier 3.] The literature PM as the initial guess, which bypasses the finder, and the literature structural parameters (Sect.~\ref{sec:regions}). This is the automatic mode of the code for a known system.
\item[Tier 4.] A dedicated run, with options tailored to the system.
\end{list}

\noindent A system counts as recovered when its membership passes the detection test of Sect.~\ref{sec:errors}, with an expected contamination below 10 percent, and its systemic PM agrees with that of the finder when the finder ran. The velocity bound of Sect.~\ref{sec:finder} applies at every tier where a distance is known, and a blind membership without a finder candidate is not a result (Sect.~\ref{sec:finder}). At tier 4, the last resort, a contamination between 10 and 20 percent is accepted and flagged. Memberships between $2$ and $3\sigma$ are marginal: listed, but kept out of the statistics. The tables give the tier of each result.

All the runs share the same configuration: the full model with the membership threshold $P \geq 0.9$, for which the mock catalogs give a purity of 90--100\% (Sect.~\ref{sec:validation}), the quasar systematics, and the \Gaia{} RVS $v_{\rm los}$ for the $v_{\rm los}$ term (no external catalog, except in the Fornax example). The literature values, used only from tier 2 onwards, are those of Simbad, of the Local Volume Database \citep[LVDB;][]{Pace2025} for the dwarf galaxies and of the \citet[2010 edition]{Harris1996} catalog for the globular clusters. Nothing else is tuned per object.

\subsection{Four examples in detail}\label{sec:examples}

\begin{figure*}
\begin{center}
\includegraphics[width=0.93\textwidth]{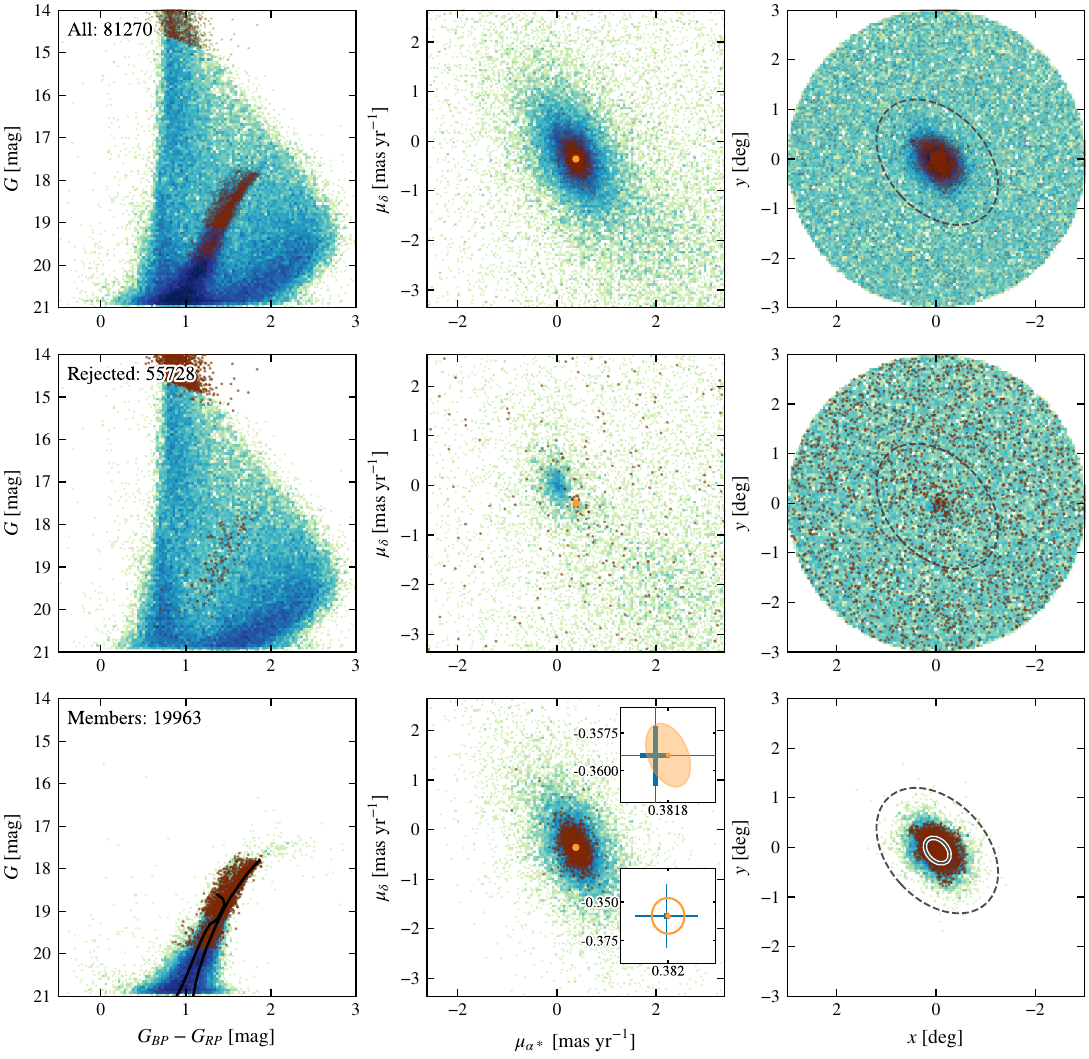}
\caption{Membership selection in the Fornax dSph galaxy. From top to bottom: all the sources of the cone, the sources rejected as non-members (the 5579 stars above the threshold that fail the quality screening are not shown), and the selected members; from left to right: CMD, PM plane and tangent-plane positions, as logarithmic density maps. The orange dot is the systemic PM of the model, the filled orange ellipse its $1\sigma$ statistical covariance and the orange outline the total one, with the quasar-based systematics added. The blue error bars are the literature value of \citet{Pace2022}: the thick bar is their statistical error and the thin bar the total one, with the \Gaia{} systematic term they list added in quadrature. The two insets zoom on them to $1.5\sigma$ of the total errors (lower) and of the statistical ones (upper). The dashed ellipse is the boundary of the object region, and the solid ellipse in the bottom panel the half-light ellipse of the fitted profile. The brick-red symbols mark the stars with a line-of-sight velocity (\Gaia{} RVS or the external catalog); those at the position of Fornax in the middle row are giants rejected on their $v_{\rm los}$ (see text). The PARSEC isochrone \citep{Bressan2012} that best fits the members at the literature distance and reddening (12.5\,Gyr, ${\rm [Fe/H]} = -1.2$) is drawn on the CMD of the members for illustration only; it is not used in the selection.}
\label{fig:fornax}
\end{center}
\end{figure*}

\begin{table*}
\caption{The four example systems, all recovered at tier 1. For each system, the first row gives the literature distance, the distance estimated by the finder (tip of the RGB for Fornax, parallax for NGC\,5466 and M67, none for Hercules), the sources in the cone, the members, the stars above the threshold that fail the quality screening, the members with a line-of-sight velocity, the detection significance, the systemic PM and $v_{\rm los,sys}$ of the members with statistical and systematic errors and the correlation $\rho$ of the two components of the total PM covariance, and the PARSEC isochrone (age and [Fe/H]) that best fits the members at the literature distance, fitted after the selection for illustration only. The second row gives the PM of the finder (Sect.~\ref{sec:detection}) with its error, and the third the literature values with their quoted errors.}
\label{tab:examples}
\centering
\footnotesize
\setlength{\tabcolsep}{3pt}
\resizebox{\textwidth}{!}{%
\begin{tabular}{l c r r r r r r r l l r l l}
\toprule
System & Tier & $D_{\rm lit}$ & $D_{\rm finder}$ & $N_{\rm cone}$ & $N_{\rm mem}$ & $N_{\rm scr}$ & $N_{v}$ & $S$ & $\pmra$ & $\pmdec$ & $\rho$ & $v_{\rm los,sys}$ & Isochrone \\
       &  & [kpc] & [kpc] &  &  &  &  &  & [$\masyr$] & [$\masyr$] &  & [$\kms$] & \\
\midrule
Fornax dSph & 1M & 147 & 138 & 81\,270 & 19\,963 & 5579 & 2155 & 2320 & $+0.382 \pm 0.002 \pm 0.011$ & $-0.359 \pm 0.002 \pm 0.011$ & $-0.02$ & $54.6 \pm 0.3$ & 12.5 Gyr, $-1.2$ \\
\quad finder & & & & & & & & & $+0.382 \pm 0.003$ & $-0.359 \pm 0.004$ & & & \\
\quad literature$^{\rm a}$ & & & & & & & & & $+0.381 \pm 0.001$ & $-0.359 \pm 0.002$ & & $55.2 \pm 0.1$$^{\rm c}$ & \\
NGC\,5466 & 1M & 16.0 & 12.1 & 21\,212 & 3\,205 & 1552 & 30 & 1051 & $-5.347 \pm 0.004 \pm 0.022$ & $-0.818 \pm 0.003 \pm 0.016$ & $-0.10$ & $108.1 \pm 0.5 \pm 0.3$ & 12.5 Gyr, $-2.1$ \\
\quad finder & & & & & & & & & $-5.351 \pm 0.010$ & $-0.827 \pm 0.008$ & & & \\
\quad literature$^{\rm b}$ & & & & & & & & & $-5.342 \pm 0.025$ & $-0.822 \pm 0.024$ & & $106.8 \pm 0.2$$^{\rm d}$ & \\
Hercules & 1M & 132 & -- & 63\,543 & 31 & 7 & 0 & 10.2 & $-0.040 \pm 0.047 \pm 0.025$ & $-0.353 \pm 0.043 \pm 0.019$ & $+0.42$ & -- & 10 Gyr, $-1.6$ \\
\quad finder & & & & & & & & & $+0.008 \pm 0.073$ & $-0.385 \pm 0.072$ & & & \\
\quad literature$^{\rm a}$ & & & & & & & & & $-0.035 \pm 0.042$ & $-0.339 \pm 0.036$ & & & \\
M67 & 1M & 0.84 & 0.84 & 119\,905 & 1\,468 & 362 & 330 & 230 & $-10.953 \pm 0.006 \pm 0.012$ & $-2.918 \pm 0.006 \pm 0.017$ & $+0.04$ & $33.89 \pm 0.18 \pm 0.30$ & 4 Gyr, $0.0$ (lit.) \\
\quad finder & & & & & & & & & $-10.958 \pm 0.008$ & $-2.914 \pm 0.007$ & & & \\
\quad literature$^{\rm e}$ & & & & & & & & & $-10.965 \pm 0.007$ & $-2.906 \pm 0.007$ & & $33.64 \pm 0.03$$^{\rm f}$ & \\
\bottomrule
\end{tabular}}\\[3pt]
\parbox{0.95\textwidth}{\footnotesize Sources of the literature values: (a) \citet{Pace2022}, statistical errors; (b) \citet{Vasiliev2021}, errors including the systematic floor; (c) \citet{Walker2009b}; (d) the compilation of \citet{Baumgardt2019}; (e) \citet{HuntReffert2023}, errors of the mean of their members; (f) \citet{Geller2015}.}
\end{table*}

Figure~\ref{fig:fornax} and, in Appendix~\ref{apx:examples}, Figs.~\ref{fig:ngc5466} to \ref{fig:m67} show, for four systems, the sources of the cone, the sources rejected as non-members, and the selected members, in the CMD, the PM plane and the sky. All four are tier-1 results: the code was given their coordinates and nothing else, with the default cone of $1.5\degree$, enlarged to $3\degree$ for Fornax and M67 by the rule of Sect.~\ref{sec:regions}. Table~\ref{tab:examples} lists the numbers, including the PM of the finder next to that of the members. The quality screening of Sect.~\ref{sec:quality} removes 22, 33, 18 and 20 percent of the stars above the membership threshold in Fornax, NGC\,5466, Hercules and M67, mostly faint or blended sources in the crowded centers (bright stars and binaries in M67). The \Gaia{} classifications of extragalactic sources, had they been used as cuts (Sect.~\ref{sec:quasars}), would have removed genuine members: among the members fainter than $G = 20$, 18 percent in NGC\,5466 and 3 of the 9 in Hercules have a DSC quasar probability above 0.5. The middle row is the test of the selection: if the model works, the rejected sample should be the field alone, with no trace of the system left in any of the three spaces. The isochrones drawn in the CMDs are PARSEC models \citep{Bressan2012} fitted to the members after the selection, by a grid search in age (5, 8, 10 and 12.5\,Gyr) and metallicity ($-2.2 \leq {\rm [Fe/H]} \leq -0.4$ in steps of 0.1\,dex) at the literature distance and reddening. They play no part in the membership and are a consistency check of the selected sequence, not a population study.

\emph{Fornax} ($D = 147\kpc$) is the most luminous of the classical dSphs and a demanding test: it is extended ($r_h \simeq 17$ arcmin), elongated ($\epsilon \simeq 0.3$), and its rich population of giants overlaps with the MW halo in the CMD. From its coordinates alone, the finder locates it at $101\sigma$ as the only candidate, with half of the excess within 13.5 arcmin and a distance of 138\,kpc from the tip of its giant branch. \getgaia{} selects 19\,963 members out of the 81\,270 sources of the cone (19\,982 without the external $v_{\rm los}$ described below), with 51 expected contaminants (0.3 percent). The rejected sample (middle row of Fig.~\ref{fig:fornax}) holds, within one half-light radius, 276 stars against 253 expected from the field, and most of the excess are the 21 stars rejected on their $v_{\rm los}$ (brick red in the figure): no trace of the galaxy is left otherwise. The members trace the RGB from its tip, at $G \simeq 17.7$, down to the magnitude limit, where the red clump of the intermediate-age population comes into reach, along a 12.5\,Gyr isochrone of ${\rm [Fe/H]} = -1.2$. The systemic PM agrees with the finder's to $0.001\masyr$ and with \citet{Pace2022}, \citet{Battaglia2022} and \citet[the first version of \getgaia{}: $(+0.380 \pm 0.001, -0.361 \pm 0.002)\masyr$]{Martinez-Garcia2021} within $0.002\masyr$ (Table~\ref{tab:examples}), and the fitted profile ($\epsilon = 0.31$, position angle $42\degree$, $r_h = 17.7$ arcmin) matches the photometric values of \citet{Battaglia2006} ($0.30 \pm 0.01$, $46.8 \pm 1.6\degree$) and those compiled in the LVDB from \citet{Wang2019} ($0.29$, $42.7\degree$, 19.9 arcmin). The 1288 quasars within $3\degree$ give a local offset of the reference frame of $(-0.003 \pm 0.010, +0.000 \pm 0.011)\masyr$ and a systematic term of $(0.011, 0.011)\masyr$, below the all-sky floor because the footprint of the galaxy averages the small-scale term. With the literature PM and structural parameters instead (tier 3), the run returns 19\,852 members and the same PM.

\begin{figure}
\begin{center}
\includegraphics[width=\columnwidth]{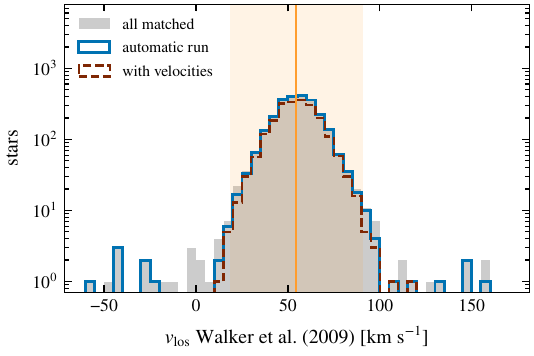}
\caption{Line-of-sight velocities as a test of the Fornax selection: distribution of the $v_{\rm los}$ of \citet{Walker2009a} for the 2611 stars of their Fornax catalog matched to the \Gaia{} sources (grey), for the stars that the tier-1 run places above the membership threshold without using any $v_{\rm los}$ (blue), and for the members of the run that uses the $v_{\rm los}$ (dashed brick red); the vertical line and the band are $v_{\rm los,sys}$ and $\pm 3\sigma_{\rm los}$ of the latter.}
\label{fig:velocities_fornax}
\end{center}
\end{figure}

Fornax also illustrates the $v_{\rm los}$ term from an external catalog. We supplied the MMFS catalog of \citet{Walker2009a}, 2633 red giant candidates with a median $v_{\rm los}$ error of $1.2\kms$, matched by position to 2611 \Gaia{} sources at the 2006 epoch of the survey. The 2586 well-measured stars with a $v_{\rm los}$ contain 2155 expected members, and the model returns $v_{\rm los,sys} = 54.6 \pm 0.3\kms$ with an intrinsic dispersion of $12.1 \pm 0.2\kms$, against the $54.8^{+0.4}_{-0.6}\kms$ of \citet{delPino2017} from these data, the $54.7 \pm 0.2\kms$ and $12.1 \pm 0.2\kms$ that \citet{Pace2021} derive from them combined with Magellan/M2FS spectroscopy, and the $11.7 \pm 0.9\kms$ of \citet{Walker2009b}. The systemic PM and the membership of the stars without a $v_{\rm los}$ do not change; among those with one, 21 former members are rejected, at a median of $150\kms$ from $v_{\rm los,sys}$, and one is added. The catalog is targeted at members and holds almost no field star, and yet the field $v_{\rm los}$ term, taken from the reflex solar motion and the 1237 RVS stars of the annulus ($29 \pm 42\kms$), is what identifies those 21 interlopers. These velocities also test the selection made without them (Fig.~\ref{fig:velocities_fornax}): the tier-1 run places 97 percent of the 2611 matched stars above the membership threshold, 97.6 percent of those that \citet{Walker2009a} classify as members, and only 1.9 percent of the stars it selects lie more than $3\sigma_{\rm los}$ from $v_{\rm los,sys}$. Those few interlopers, which no astrometric or photometric criterion can separate from the giants of Fornax, are what the $v_{\rm los}$ term removes.

\emph{NGC\,5466} ($D = 16.0\kpc$; Fig.~\ref{fig:ngc5466}) is a loose, metal-poor globular cluster of the outer halo at $b = +74\degree$, which stresses the model in the opposite way to Fornax: a compact system with a small intrinsic PM dispersion and a large PM. The finder locates it at $44\sigma$, with half of the excess within 3.1 arcmin and a parallax distance of 12\,kpc, and the code selects 3205 members out of the 21\,212 sources of the cone, with 5 expected contaminants. The rejected sample is the field alone: within the half-light radius it holds 34 stars against 19 expected, the excess being stars of intermediate probability at the faint end of the main sequence, where the astrometric errors exceed $1\masyr$. The members trace the RGB, the HB and the main sequence down to $G \simeq 21$, along a 12.5\,Gyr isochrone of ${\rm [Fe/H]} = -2.1$, close to the $-2.0$ of the \citet{Harris1996} catalog. The systemic PM agrees with the finder's within $0.01\masyr$ and with VB21 within $0.006\masyr$. The cluster is bright enough for the \Gaia{} RVS: 30 members have a $v_{\rm los}$, and the $v_{\rm los}$ term gives $v_{\rm los,sys} = 108.1 \pm 0.5 \pm 0.3\kms$ (statistical and zero-point errors) with an intrinsic dispersion of $0.9 \pm 0.7\kms$, within $2\sigma$ of the $106.8 \pm 0.2\kms$ of \citet{Baumgardt2019}. With the literature PM and half-light radius instead (tier 3), the run returns 3161 members and the same PM.

\emph{Hercules} ($D = 132\kpc$, $M_V = -5.8$; Fig.~\ref{fig:hercules}) represents the ultra-faint regime: its RGB starts at $G \simeq 18$, and only 31 of the 63\,543 sources of the cone are members. The finder locates it at $5.2\sigma$, as the only candidate, from 77 stars at the peak, with half of the excess within 4 arcmin; its PM, $(+0.01 \pm 0.07, -0.39 \pm 0.07)\masyr$, agrees with that of the members within the errors, and no distance can be estimated from so few giants. The preliminary profile wanders to a half-light radius of $0.25\degree$, is rejected against the size of the finder and replaced by a circle, which the EM refines to 5 arcmin. The rejected sample is indistinguishable from the field, and the members form a narrow sequence along a 10\,Gyr, ${\rm [Fe/H]} = -1.6$ isochrone, clustered in PM and on the sky inside the half-light ellipse. The isochrone is more metal-rich than the spectroscopic metallicity of the galaxy, $-2.6$ \citep[as compiled in the LVDB from][]{Geha2026}, which shows the limits of a fit of this kind: over the two magnitudes of RGB that the members span, the most metal-poor isochrone of our grid is bluer by only $0.04$--$0.12$\,mag, less than their color errors, so three bright stars decide the result. The selection, which uses no isochrone, is unaffected: the detection is at $10.2\sigma$ with 1.6 contaminants expected, and the systemic PM agrees with \citet{Pace2022} within $0.014\masyr$. Here the three terms of the model work together. With PM errors of $0.3$--$1\masyr$ none of these stars reaches a high probability from the astrometry alone; it is their concentration on the sky and in the CMD that selects them, while the field-based test guards against a spurious concentration. With the literature PM and the LVDB shape instead (tier 3), the run returns 32 members at $7.9\sigma$ and a PM within $0.014\masyr$ of this one.

\emph{M67} (NGC\,2682; $D = 0.84\kpc$; Fig.~\ref{fig:m67}) is an open cluster and the opposite case in every respect: a parallax of $1.2\mas$, a PM of $11\masyr$, a main sequence instead of a giant branch, and a disk field at $b = +32\degree$. From its coordinates alone, in a cone of 119\,905 sources with no PM pre-selection, the finder locates the cluster at $30\sigma$ with a parallax of $1.19\mas$, hence 840\,pc, and the code selects 1468 members with 17 expected contaminants. They trace the main sequence from the turnoff at $G \simeq 12.5$ down to $G = 21$, the subgiant and red giant branches, the red clump and the binary sequence; a 4\,Gyr isochrone of solar metallicity, drawn at the literature values, follows them. The systemic PM agrees with the finder's within $0.005\masyr$ and with that of \citet[HR23]{HuntReffert2023} within $0.012\masyr$ in each component, the intrinsic dispersion, $(0.20, 0.19)\masyr$, with their $(0.185, 0.187)\masyr$, and the systemic parallax, $1.191 \pm 0.002 \pm 0.021\mas$, places the cluster at 840\,pc, against their 837\,pc. Of the 1842 HR23 members in the cone, 1687 have $P \geq 0.9$ and 103 fall below 0.5, bright stars (median $G = 15.5$) in the outskirts of the cluster; 350 of their stars fail the quality screening, mostly bright stars and binaries with a large RUWE. The 330 members with a \Gaia{} RVS velocity give $v_{\rm los,sys} = 33.89 \pm 0.18 \pm 0.30\kms$, against the $33.64 \pm 0.03\kms$ of \citet{Geller2015} from four decades of ground-based spectroscopy, with an intrinsic dispersion of $1.2 \pm 0.3\kms$ that includes the unresolved binaries. With the PM of Simbad as the guess (tier 3), in a cone of $1.8\degree$ set by the Simbad size, the run returns 1405 members and the same PM.

\subsection{The dwarf galaxies of the Local Group}\label{sec:localgroup}

\begin{figure}
\begin{center}
\includegraphics[width=\columnwidth]{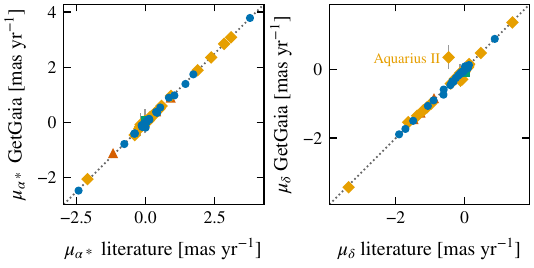}
\caption{Systemic PMs of the Local Group dwarf galaxies recovered by \getgaia{}, $\pmra$ (left) and $\pmdec$ (right), against the literature values compiled in the LVDB \citep[][mostly from \citealt{Pace2022}, \citealt{Battaglia2022} and \citealt{Bennet2024}]{Pace2025}; the error bars include the statistical and the quasar-based systematic uncertainties. Circles, squares, diamonds and triangles are the systems recovered at tiers 1 to 4 (Sect.~\ref{sec:application}); Aquarius II, the one system that stands out, is labeled.}
\label{fig:localgroup}
\end{center}
\end{figure}

\begin{figure}
\begin{center}
\includegraphics[width=\columnwidth]{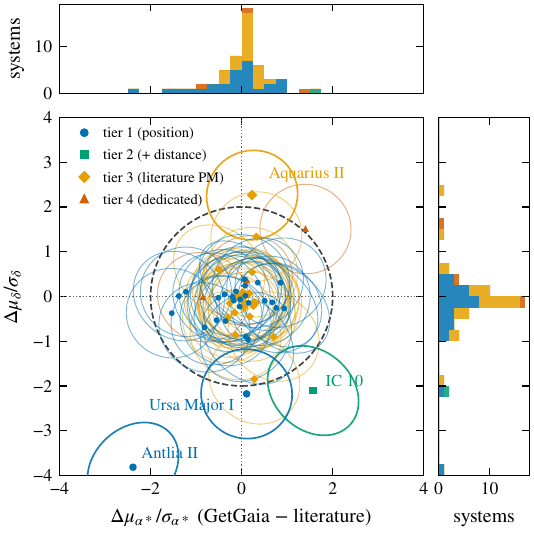}
\caption{Differences between our systemic PMs of the Local Group dwarfs and the literature values, in units of the combined errors, with the $1\sigma$ ellipse of the combined covariance of each system (our full covariance, statistical and systematic, plus the literature errors); in these units the ellipses have unit size and their orientation shows the correlation between the two components, and the dashed circle marks the $2\sigma$ bound. The histograms give the distribution of each component, stacked by tier (values beyond the range are placed in the outer bins). Symbols as in Fig.~\ref{fig:localgroup}; the systems that deviate by more than $2\sigma$ in either component or with the full covariance are labeled, at the edge of the panel when they fall outside it.}
\label{fig:localgroup_pulls}
\end{center}
\end{figure}

To test the code over a wide range of conditions (distance, luminosity, surface brightness, Galactic latitude), we ran it on the 103 known dwarf galaxies of the Local Group within 850\,kpc of the Sun. The Magellanic Clouds, the Sagittarius dSph, M\,31 and M\,33, which extend over several degrees and require dedicated settings, were excluded. The sample deliberately extends far beyond the reach of \Gaia{} for an old stellar population at $G = 21$, as we discuss below.

At tier 1 the distance came from the onset of the RGB (Sect.~\ref{sec:finder}) for 18 of the 29 systems recovered; at tier 2 it was that of the LVDB. At tier 3 the literature PM \citep[Simbad, else][]{McConnachie2020b, Pace2022} replaced the finder, and the structural parameters of the LVDB, including its tables of recently discovered and ambiguous systems, were supplied as the literature shape of Sect.~\ref{sec:regions}. No dwarf galaxy has three members bright enough for the \Gaia{} RVS, so the $v_{\rm los}$ term acts only on the bright foreground stars of each cone, and the results are the same, to within $0.01\masyr$, as without it.

The detection criterion, an expected membership of at least five stars at $3\sigma$ over the field, is deliberately stricter than common practice. Systemic PMs of ultra-faint dwarfs are routinely published from about four candidate members, not always confirmed by spectroscopy \citep[e.g. Boötes IV, Horologium II, Pegasus III and Tucana V in][]{Pace2022}. Three systems reach $2$--$3\sigma$ at some tier without passing it, Boötes IV, Tucana IV and NGC\,185: they are listed as marginal in Table~\ref{tab:lg} and excluded from the statistics below.

\begin{table}
\caption{Dwarf galaxies recovered at each tier (Sect.~\ref{sec:application}): number of systems in each group, recovered at tiers 1 to 4, recovered in total, marginal ($2$--$3\sigma$) and not recovered. The galaxies within 300\,kpc are split at $M_V = -8$ into the classical dSphs and the ultra-faint dwarfs (UFDs).}
\label{tab:tiers}
\centering
\scriptsize
\setlength{\tabcolsep}{2.5pt}
\begin{tabular}{l r r r r r r r r}
\toprule
 & $N$ & Tier 1 & Tier 2 & Tier 3 & Tier 4 & Rec. & Marg. & No \\
\midrule
dSphs ($M_V \leq -8$) & 11 & 10 & 0 & 1 & 0 & 11 & 0 & 0 \\
UFDs ($M_V > -8$) & 51 & 12 & 0 & 20 & 3 & 35 & 2 & 16 \\
Dwarfs beyond 300\,kpc & 41 & 7 & 1 & 1 & 0 & 9 & 1 & 32 \\
\midrule
All dwarfs & 103 & 29 & 1 & 22 & 3 & 55 & 3 & 48 \\
\bottomrule

\end{tabular}
\end{table}

Table~\ref{tab:tiers} summarizes the recoveries by tier and Table~\ref{tab:lg} (Appendix~\ref{apx:lg}) lists them: 55 galaxies, 29 from the position alone, one (IC\,10) with the literature distance, 22 with the literature PM and three with a dedicated run. The finder found a candidate in 55 of the 103 cones at tier 1, but 25 of them were groups of foreground stars, 16 in the fields of M\,31 satellites, which the velocity bound discards at tier 2 once the distance is known; seven more memberships, seeded by a genuine candidate, absorbed 14 to 41 percent of field stars sharing the PM of the galaxy. The 22 systems that need the literature PM are ultra-faint dwarfs whose few stars do not stand out in the PM space of their cone, plus Leo T at 420\,kpc. Carina II is the clearest case: embedded in the outer halo of the LMC, whose stars share its PM, it shows no excess in PM space, and it is the CMD term, through the field-based test of Sect.~\ref{sec:errors}, that separates its old, metal-poor giants from the LMC population once the PM is given. Boötes III, Tucana II and Grus II stay contaminated even then, and their dedicated runs (the literature shape, the object region at two half-light radii, the dispersion cap of a $10\kms$ system and $P \geq 0.95$) bring them to 19, 13 and 7 percent, the first two with the warning of Table~\ref{tab:lg}.

The results agree with the literature. Of the 55 galaxies, 53 have a PM in the LVDB, mostly from \citet{Pace2022}, \citet{Battaglia2022} and \citet{Bennet2024}, and ours agree with it within $0.1\masyr$ for 47 and within $0.3\masyr$ for 52 (Fig.~\ref{fig:localgroup}), with a median difference of $0.032\masyr$ and a median deviation of $0.37$ combined errors (Fig.~\ref{fig:localgroup_pulls}), equally well at every tier ($0.020\masyr$ for the 27 tier-1 systems with a literature PM, $0.033\masyr$ for the 22 tier-3 ones). The 13 galaxies that \citet{Martinez-Garcia2021} measured with the first version of \getgaia{}, from the sequential cuts of Sect.~\ref{sec:intro}, are all recovered here, within $0.05\masyr$ of their values for ten of them and within $0.15\masyr$ ($2.0\sigma$, Reticulum II) for all, with a median difference of $0.020\masyr$. No system deviates by more than $3\sigma$ from the LVDB; Antlia II, discussed below, is the only one beyond $2\sigma$ with the full covariance, and Ursa Major I, IC\,10 and Aquarius II exceed $2\sigma$ in $\pmdec$ alone, the last with a difference of $0.82\masyr$ that rests on three stars above the threshold. The correlation between the two PM components has a median absolute value of 0.14 and reaches 0.48 (Sculptor). The 55 systems hold 51\,648 members in total, with an expected contamination of 1.1 percent from the probabilities and 2.8 percent from the field-based estimate. For comparison, \citet{Battaglia2022} obtained reliable systemic PMs for 66 of 74 Local Group dwarfs with a similar mixture model informed by spectroscopic members, literature structural parameters and CMD filters; without spectroscopy, \getgaia{} measures 46 of the 62 systems of our sample within 300\,kpc, 27 of them from the position alone, plus nine galaxies beyond.

The recoveries include every system within 300\,kpc brighter than $M_V = -5.4$, ultra-faint dwarfs down to $M_V = -0.8$ (Draco II, at 22\,kpc) and to $M_V = -4.2$ at 183\,kpc (Columba I), and nine galaxies at 366--824\,kpc. The distant ones are of three kinds. Leo T and Eridanus II, at $M_V = -8.0$ and $-7.1$, are the faintest old systems recovered beyond 300\,kpc, through their upper RGB. Phoenix, NGC\,6822, IC\,1613, Leo\,A and IC\,10 are star-forming galaxies, recovered through their young supergiants and giants, which reach $G = 21$ where their old populations do not. M\,32 and NGC\,205 are the bright companions of M\,31, recovered from the position alone with 6754 and 6176 members, but the population that the finder locates is the disk of M\,31 behind them, and the PMs returned, $(+0.029, -0.041)$ and $(+0.022, -0.045)\masyr$, are those of M\,31 \citep[$(+0.049 \pm 0.011, -0.037 \pm 0.008)\masyr$ from \Gaia{} EDR3,][consistent with the HST and \Gaia{} DR2 measurements of \citealt{Sohn2012}, \citealt{vanderMarel2012} and \citealt{vanderMarel2019}]{Salomon2021}. At the other end, Ursa Major III/UNIONS 1 \citep{Smith2024}, the faintest known satellite, at 10\,kpc with a half-light radius of 3\,pc, has no candidate in PM space and, with the literature PM, 4 members, below our criterion.

The 48 systems that no tier recovers are, with few exceptions, beyond the reach of \Gaia{} rather than failures of the method. For an old population the HB ($M_G \simeq +0.4$) is lost beyond 130\,kpc and the tip of the RGB ($M_G \simeq -3.2$) beyond 680\,kpc, and scaled from the mock of Sect.~\ref{sec:validation}, five stars brighter than $G = 21$ require $M_V \lesssim -3$ at 126\,kpc, $-4.5$ at 200\,kpc and $-5.5$ at 316\,kpc. Accordingly, with the literature PM, 45 of the 48 have fewer than five members and none exceeds $2.5\sigma$. The 32 beyond 300\,kpc return at most a few stars sharing the expected astrometry, and the 16 within 300\,kpc are all fainter than $M_V = -5.3$: three have no member at all, and the others return between one and nine members below the criterion, like Tucana IV (9 members at $2.5\sigma$, within $0.3\masyr$ of the literature PM). Pisces II and Indus II are blind runs without a candidate (Sect.~\ref{sec:finder}) and are therefore not reported.

The spatial term also returns the half-light radius, the ellipticity and the position angle of the 28 recovered dwarfs whose profile was fitted freely, compared with the photometric values of the LVDB in Appendix~\ref{apx:structure}: the position angles agree within $7\degree$ (median) and the ellipticities within 0.01, while the half-light radii come out 21 percent larger, a systematic difference discussed there.

Antlia II, the one system beyond $2\sigma$, is recovered from the position alone with 243 members at $10.7\sigma$ and 9 percent contamination, and its PM differs from that of \citet{Pace2022} by $(-0.042, -0.064)\masyr$ ($2.9\sigma$ with the full covariance); its tier-3 run gives the same PM to within $0.01\masyr$. It is the most extended galaxy of the sample ($r_h = 1.7\degree$, $\epsilon = 0.6$), under tidal disruption \citep{Ji2021}, and even with its cone grown to $3\degree$ its object region fills most of it, so that the control annulus contains its outer stars. A dedicated run with a $4\degree$ cone, the object region at $2.6\degree$, the LVDB shape, the dispersion cap of a $5\kms$ system and $P \geq 0.95$ brings the difference down to $(-0.021, -0.005)\masyr$ (103 members, $1.7\sigma$ at most), with the contamination at the 10 percent limit. The stars of a disrupting galaxy spread over tens of square degrees cannot be separated cleanly from a field that shares their PM to within $0.1\masyr$, and such a system is better served by spectroscopic members \citep{Ji2021}. The contaminated memberships of tier 1 tell the same story at a smaller scale: Segue 1, which lies on the Sagittarius stream \citep{Niederste-Ostholt2009}, and Tucana III, with tidal tails of its own, need the literature PM and shape to keep the stars that share their PM in the field component, and then agree with the literature within $0.1\masyr$.

\subsection{Milky Way globular clusters}\label{sec:gcs}

\begin{figure}
\begin{center}
\includegraphics[width=\columnwidth]{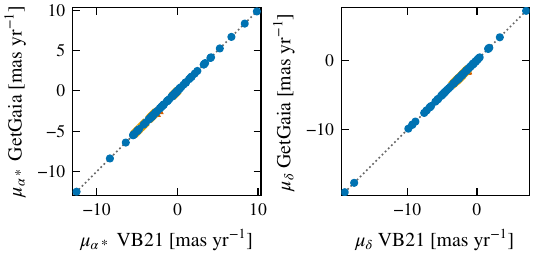}
\caption{Same as Fig.~\ref{fig:localgroup} for the 93 Galactic globular clusters recovered, against \citet[VB21]{Vasiliev2021}: circles for the 86 recovered from the position alone, diamonds for the five that need the literature PM and triangles for the two dedicated runs (Sect.~\ref{sec:gcs}).}
\label{fig:gcs}
\end{center}
\end{figure}

\begin{figure}
\begin{center}
\includegraphics[width=\columnwidth]{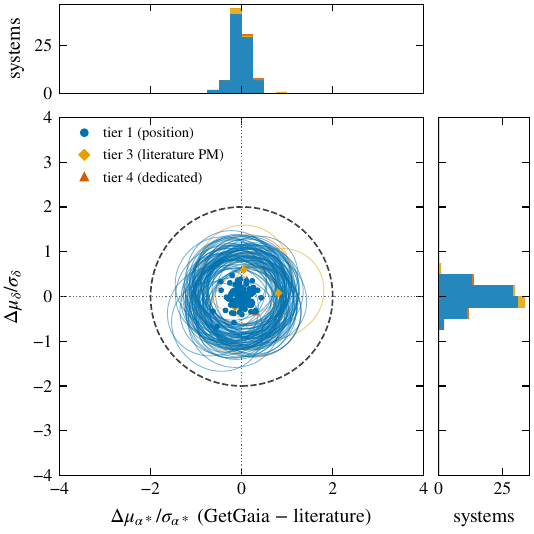}
\caption{Same as Fig.~\ref{fig:localgroup_pulls} for the globular clusters, against VB21.}
\label{fig:gcs_pulls}
\end{center}
\end{figure}

The same sequence was applied to the 95 Galactic globular clusters of the \citet[2010 edition]{Harris1996} catalog that lie outside the bulge and the inner disk; the 56 clusters at $|b| < 5\degree$, or otherwise known to sit in heavily reddened, crowded fields, require dedicated settings and were excluded. The literature values are those of the \citet{Harris1996} catalog: its distance at tier 2, and at tier 3 the PM listed by Simbad and its half-light radius; the catalog gives no ellipticity or position angle, so the literature shape is a circle. Many clusters have giants within reach of the \Gaia{} RVS, so the $v_{\rm los}$ term acts on their own stars.

The tiers add little here: the code recovers 93 clusters (Fig.~\ref{fig:gcs} and Tables~\ref{tab:gcs_a} and \ref{tab:gcs_b} in Appendix~\ref{apx:gcs}), 86 of them from the position alone, five with the literature PM and two with a dedicated run, and tier 2 recovers none. They hold $4.4\times10^5$ members in total (from 6 in AM\,4 to 74\,343 in $\omega$ Centauri), 1.5 times the number of stars used by \citet[VB21]{Vasiliev2021}, who combined a similar mixture model with the structural parameters of each cluster. The systemic PMs of VB21, which include a systematic error floor of $0.025\masyr$, agree with ours within $0.02\masyr$ for 85 clusters and within $0.08\masyr$ for all 93, with a median difference of $0.009\masyr$ and a median deviation of $0.18$ combined errors (Fig.~\ref{fig:gcs_pulls}); none deviates by more than $2\sigma$. The correlation between the PM components has a median absolute value of 0.09 and reaches 0.67 (NGC\,7006), the systemic parallaxes agree with VB21 within a median absolute deviation of $0.003\mas$, and the expected contamination is 0.7 percent overall. The two clusters not recovered, Eridanus and Crater, have at most four stars above the threshold at every tier (VB21 have 44 and 15 stars for them); Crater reaches $2.6\sigma$ with the literature PM and is listed as marginal. AM\,4, with six members at $5\sigma$, is the sparsest system recovered and the one farthest from VB21 ($0.08\masyr$, within $1\sigma$ of their 8-star value).

The special cases are the clusters in dense fields and those inside the Sagittarius dwarf. Three clusters at $|b| < 11\degree$, in fields of $10^5$--$10^7$ stars, need the literature PM: for NGC\,6535 the finder locks on the disk population that dominates the PM space of its cone, and for NGC\,5927 and NGC\,6284 the position-only runs, whose cones grow to $3\degree$ and seven million sources, exceeded the memory of our machine (32 GB). With the literature PM and half-light radius the three are recovered within $0.03\masyr$ of VB21, as are NGC\,6266 and NGC\,6717, in similar fields, from the position alone; in such fields the spatial term is what keeps the object component from absorbing the disk population. NGC\,6715 (M\,54), in the nucleus of the Sagittarius dwarf, and Terzan 7 and Arp 2, embedded in it, absorb 10 to 35 percent of stars of the dwarf from the position alone, which share their PM and CMD. The literature PM and half-light radius bring Terzan 7 to 5 percent contamination and within $0.04\masyr$ of VB21, but leave 11 and 14 percent in M\,54 and Arp 2, which end at tier 4. A dedicated run with the object/field boundary at three to four half-light radii, instead of five, and the dispersion cap of a $10$--$15\kms$ cluster gives systemic PMs within $0.02\masyr$ of VB21 for both, with 4 percent contamination for Arp 2; for M\,54, even with $P \geq 0.95$, 11 percent of the 128 members are still expected to be stars of the Sagittarius nucleus, and it carries the warning of Table~\ref{tab:gcs_b}.

\begin{figure}
\begin{center}
\includegraphics[width=\columnwidth]{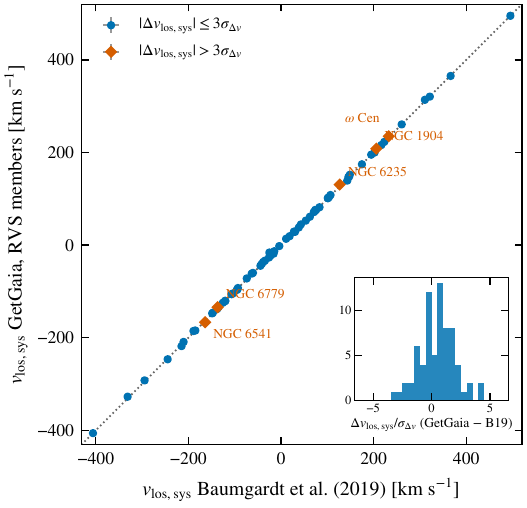}
\caption{Line-of-sight velocities as a test of the cluster selection: $v_{\rm los,sys}$ of the 71 globular clusters with at least three members in the \Gaia{} RVS, from those members alone, against the compilation of \citet{Baumgardt2019}; the inset shows the distribution of the differences $\Delta v_{\rm los,sys}$ in units of their combined error $\sigma_{\Delta v}$, and the orange diamonds are the five clusters beyond $3\sigma_{\Delta v}$, discussed in the text.}
\label{fig:velocities}
\end{center}
\end{figure}

The $v_{\rm los}$ term returns a systemic line-of-sight velocity and an intrinsic dispersion from the \Gaia{} data alone for the 71 recovered clusters with at least three members with a $v_{\rm los}$ (38 in the median, from 4 to 908 in 47 Tucanae; Fig.~\ref{fig:velocities}), with a median total error of $0.9\kms$. Compared with the compilation of \citet{Baumgardt2019}, built from ground-based spectroscopy with a median error of $0.3\kms$, the differences have a median of $+0.5\kms$ and a median absolute deviation of $1.0\kms$, and 58 of the 71 agree within $2\sigma$. The offset has the size and sign of the known zero-point trend of the RVS $v_{\rm los}$ of faint stars \citep{Katz2023}, which the code does not correct. Five clusters differ by more than $3\sigma$: $\omega$ Centauri, whose rotation and $10\kms$ dispersion are not a single Gaussian, and NGC\,1904, NGC\,6235, NGC\,6541 and NGC\,6779, with offsets of $2$--$4\kms$. The intrinsic dispersions, $0.5$--$17\kms$ with a median of $3\kms$, refer to the bright giants in the outskirts that \Gaia{} resolves, with the zero-point trend inflating them for the faintest stars, and are a by-product rather than a dynamical measurement.

\subsection{The uncertainties compared with the literature}\label{sec:uncertainties}

\begin{figure}
\begin{center}
\includegraphics[width=\columnwidth]{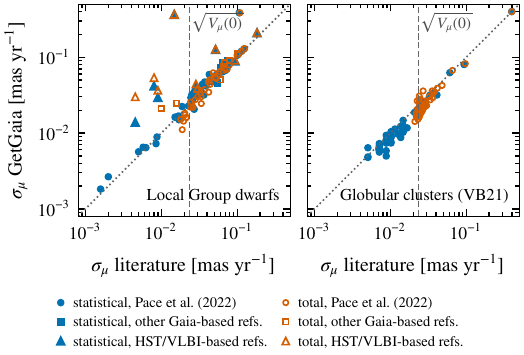}
\caption{Uncertainties of the systemic PMs (quadratic mean of the two components) against the literature ones, for the Local Group dwarfs (left) and the globular clusters (right). Filled symbols: statistical errors on both axes, ours against those of \citet{Pace2022} (circles), of other \Gaia{}-based works \citep[squares;][]{Cerny2021, Cerny2023, Heiger2024, Chiu2026}, of works that involve \emph{HST} astrometry or VLBI masers \citep[triangles;][]{Bennet2024, Bennet2025, Julio2024, Warfield2026, Brunthaler2007}, and against the VB21 errors with their systematic floor of $0.025\masyr$ removed in quadrature (clusters whose VB21 error equals the floor are omitted). Open symbols: total errors on both axes, ours with the quasar-based systematics against the literature statistical errors with the systematic term of \citet{Pace2022} added in quadrature, against the errors of \citet{Battaglia2022}, which already include the systematics (squares), against the quoted errors of the \emph{HST} and VLBI works, which are not limited by the \Gaia{} frame, and against the VB21 totals. The dashed vertical line marks $\sqrt{V_\mu(0)} = 0.023\masyr$, the zero-separation value of the covariance function of Eq.~\ref{eq:kernel}: the systematic floor that the all-sky prescriptions impose on a compact system.}
\label{fig:uncertainties}
\end{center}
\end{figure}

The insets of Figs.~\ref{fig:fornax} and \ref{fig:ngc5466} invite a comparison of the error bars derived here and those from the literature. Fig.~\ref{fig:uncertainties} shows this comparison for the whole sample. The statistical errors of \getgaia{} are the same as those of the literature. The ratio to the statistical errors of \citet{Pace2022} has a median of 1.06 (0.99--1.22 between the 16th and 84th percentiles) over the 36 dwarfs in common, and the ratio to the statistical part of the VB21 errors has a median of 0.88 (0.76--0.99), consistent with our 1.5 times larger member samples. This is expected. Both the literature and our model weight each star by its own covariance, so that for a given set of members the statistical precision is fixed by the \Gaia{} errors and the number of stars, and the inverse Fisher information of Sect.~\ref{sec:errors} simply recovers it. The two outliers are Leo I and Leo II, whose literature PMs \citep{Bennet2024} come from \textsc{GaiaHub} \citep{delPino2022}, which combines \Gaia{} with \emph{HST} astrometry and is three times more precise than \Gaia{} alone at those distances.

The second thing we notice is that the systematic term is of the same order as the one adopted in the literature, but it is measured rather than assumed. \citet{Pace2022} and VB21 take it from the all-sky covariance function of the \Gaia{} reference frame \citep{Lindegren2021a, Vasiliev2021}, which gives $0.016$--$0.023\masyr$ at the angular scale of a dwarf galaxy and a floor of $0.025\masyr$ for the clusters. \getgaia{} measures the local offset of the frame on the quasars within $3\degree$ of each system (Sect.~\ref{sec:errors}; a median of 1402 quasars for the dwarfs and 547 for the clusters, whose cones lie at lower latitude). The resulting term has a median of $0.018\masyr$ for the dwarfs ($0.014$--$0.022\masyr$) and $0.021\masyr$ for the clusters ($0.019$--$0.023\masyr$). Compared on the same footing (open symbols in Fig.~\ref{fig:uncertainties}), the total errors match those of the literature. The ratio to the total errors of \citet{Pace2022} has a median of 1.00 (0.83--1.09). The ratio to the VB21 totals has a median of 0.85, smaller because the floor of VB21 is the small-separation limit of the covariance function, which they regard as the irreducible systematic error of a compact system, whereas the footprint of a cluster that \Gaia{} resolves over several arcminutes, and the quasars inside it, average part of the small-scale term; VB21 themselves note that their uncertainty may be smaller for clusters that span more than a fraction of a degree. The local offsets themselves are small. Their median modulus is $0.010\masyr$, their median uncertainty $0.016$--$0.019\masyr$, and only two of the 147 systems have an offset beyond $2\sigma$ in either component, fewer than chance would give. The reduced $\chi^2$ of the quasar residuals has a median of 0.97 and never exceeds 1.5, so that the inflated catalog errors of the quasars (Sect.~\ref{sec:quality}) are well calibrated and the covariance of Eq.~\ref{eq:kernel} describes the fields of this sample. The pulls of Figs.~\ref{fig:localgroup_pulls} and \ref{fig:gcs_pulls}, with medians of 0.37 and 0.19 and only Antlia II beyond $2\sigma$ with the full covariance, show that the budget is, if anything, still conservative. For systems with thousands of members the statistical error is irrelevant and the systemic PM is limited by the reference frame; for the ultra-faint dwarfs, with a few tens of members, the statistical error dominates.

\section{Conclusions}\label{sec:conclusions}

We have presented \getgaia{} 2.0, a complete rewrite of the tool used in \citet{delPino2021} and \citet{Martinez-Garcia2021} to select the member stars of stellar systems in the \Gaia{} catalog. The sequential cuts of the original version are replaced by a single probabilistic model, fitted by expectation maximization, that combines the astrometry, the CMD, the sky position and, when available, the line-of-sight velocity of every star, each with its own uncertainties and correlations. The code finds the system in PM space from its coordinates alone, or starts from the literature values when they are supplied, measures the local offset of the \Gaia{} reference frame on the quasars around it, and returns the systemic motion with its full covariance, the structural parameters of the system and the membership probability of every star, together with a detection significance and a contamination estimate that do not depend on the calibration of the probabilities. All the release-dependent quantities are isolated in a registry, so that the code is ready for \Gaia{} DR4. Our main results are the following.
\begin{itemize}
\item On mock catalogs built from a metal-poor isochrone with the \Gaia{} DR3 error models and a real \Gaia{} field, the model recovers more than 94\% of the members with a purity above 92\% at $P \geq 0.5$, and of 98--100\% at the threshold $P \geq 0.9$ used throughout the paper, for systems between 50 and 316\,kpc and field densities up to thirty times that of a high-latitude field, with unbiased systemic PMs. When the system sits on the PM of the field, the purity at $P \geq 0.95$ stays above 97 percent in the Fornax field and above 90 percent in a field ten times denser, at the cost of the completeness, which falls to 50--60 percent in the densest cells, and of a PM shift of the size of the systematic term. With the DR4 error models the purity at $P \geq 0.5$ rises to 97--100\% and the PM errors shrink by a factor of about two.
\item Applied to the Fornax dSph, the GC NGC\,5466, the ultra-faint dwarf Hercules and the open cluster M67 from their coordinates alone, the code locates the four systems in PM space, Hercules included (31 stars among 63\,000), and recovers in a few minutes systemic PMs within $0.014\masyr$ of the best determinations in the literature, structural parameters consistent with deep photometry for Fornax, systemic $v_{\rm los}$ consistent with spectroscopy, and for M67 a membership that holds 92 percent of the stars of the dedicated catalog of \citet{HuntReffert2023} at $P \geq 0.9$. With the literature PM and shape instead, the runs return the same PMs.
\item Run on the 103 dwarf galaxies of the Local Group within 850\,kpc, with no spectroscopic or isochrone input, it measures 55 systems: 29 from the position alone, one more with the literature distance, 22 with the literature PM and three with a dedicated run. Their systemic PMs agree with the published values within $0.1\masyr$ for 47 of them (median difference $0.032\masyr$, median deviation $0.37\sigma$), equally well at every tier, and none deviates by more than $3\sigma$; for the 13 galaxies of \citet{Martinez-Garcia2021} the median difference is $0.020\masyr$. Of the 48 systems not recovered, 45 have fewer than five stars brighter than $G = 21$ and fail our detection criterion, an expected membership of at least five stars at $3\sigma$ over the field.
\item On 95 Galactic globular clusters outside the bulge, the same sequence recovers 93, 86 of them from the position alone, with systemic PMs within $0.08\masyr$ of \citet{Vasiliev2021} (median difference $0.009\masyr$, none beyond $2\sigma$) and 1.5 times as many members, and measures the systemic line-of-sight velocities of 71 of them from the \Gaia{} RVS alone, within $1\kms$ of the compilation of \citet{Baumgardt2019} in the median. The two clusters not recovered have fewer than five stars in \Gaia{}; five need the literature PM, three of them in dense disk fields, and M\,54 and Arp 2, embedded in the Sagittarius dwarf, a dedicated run.
\item For the 28 dwarfs whose profile is fitted freely, the position angles and ellipticities agree with the photometric ones within $2\sigma$ in about 70 percent of the cases, and the half-light radii come out 21 percent larger in the median, since \Gaia{} traces the giants of each system rather than its full light.
\item The local offsets of the reference frame, measured on the quasars of each field, have a median modulus of $0.010\masyr$. The resulting systematic term of the systemic PMs is $0.02\masyr$ for both samples, the level of the all-sky floors adopted in the literature. The statistical errors equal those of dedicated studies.
\end{itemize}
The model makes no assumption on the nature of the system. The same configuration that isolates a dwarf galaxy at 800\,kpc isolates an open cluster at 840\,pc, and \getgaia{} can detect and analyze any comoving stellar system within the reach of \Gaia{}: globular and open clusters, dwarf galaxies, moving groups, and the dense parts of a stream. A field density that varies across the cone is the main limitation of the spatial term, and a gradient term for the field is planned. Since all the release-dependent quantities live in the registry, running the code on \Gaia{} DR4 will require only its calibration values, and the mock grid shows what to expect: purities above 97 percent and PM errors halved. The code is free, documented and tested, and we plan to maintain it through DR4 and the releases that follow.

\begin{acknowledgements}
A.~del~Pino acknowledges financial support from the \emph{Severo Ochoa} grant CEX2021-001131-S (MICIU/AEI/10.13039/501100011033), from the Spanish Ministry of Science and Innovation project PID2024-155572NB-C22 (MICIU/AEI/10.13039/501100011033, FEDER, EU), from the RyC-MAX grant 20245MAX008 (CSIC), from the Ram\'on y Cajal fellowship RYC2022-038448-I (MICIU/AEI/10.13039/501100011033, co-funded by the European Social Fund Plus), and from grant CNS2025-166660. This work has made use of data from the European Space Agency (ESA) mission \Gaia{} (\url{https://www.cosmos.esa.int/gaia}), processed by the \Gaia{} Data Processing and Analysis Consortium (DPAC, \url{https://www.cosmos.esa.int/web/gaia/dpac/consortium}). Funding for the DPAC has been provided by national institutions, in particular the institutions participating in the \Gaia{} Multilateral Agreement. This research has made use of the SIMBAD database, operated at CDS, Strasbourg, France, and of \textsc{astropy} \citep{astropy2022}, \textsc{astroquery} \citep{Ginsburg2019}, \textsc{numpy} \citep{Harris2020}, \textsc{scipy} \citep{Virtanen2020}, \textsc{pandas} \citep{McKinney2010} and \textsc{matplotlib} \citep{Hunter2007}. The code of \getgaia{} 2.0 was developed with the assistance of a large language model, used as a programming assistant under the direction of the author; all the algorithms were specified, reviewed and validated by the author through the tests described in Sect.~\ref{sec:validation} and the comparisons with the literature of Sect.~\ref{sec:application}.
\end{acknowledgements}

\section*{Data availability}
\getgaia{} 2.0 will be made publicly available upon acceptance of the paper at \url{https://github.com/AndresdPM/GetGaia2}, under a BSD 2-clause license, together with the catalogs of members and the summary files of all the systems analyzed here. The tables of Appendices~\ref{apx:lg}, \ref{apx:gcs} and \ref{apx:structure} will also be available at the CDS via anonymous ftp to \url{cdsarc.cds.unistra.fr} (130.79.128.5) or via \url{https://cdsarc.cds.unistra.fr/cgi-bin/qcat?J/A+A/}.

\bibliographystyle{aa}
\bibliography{getgaia2}

\appendix
\section{The mock grid with the \Gaia{} DR4 error models}\label{apx:dr4}

\begin{figure}
\begin{center}
\includegraphics[width=\columnwidth]{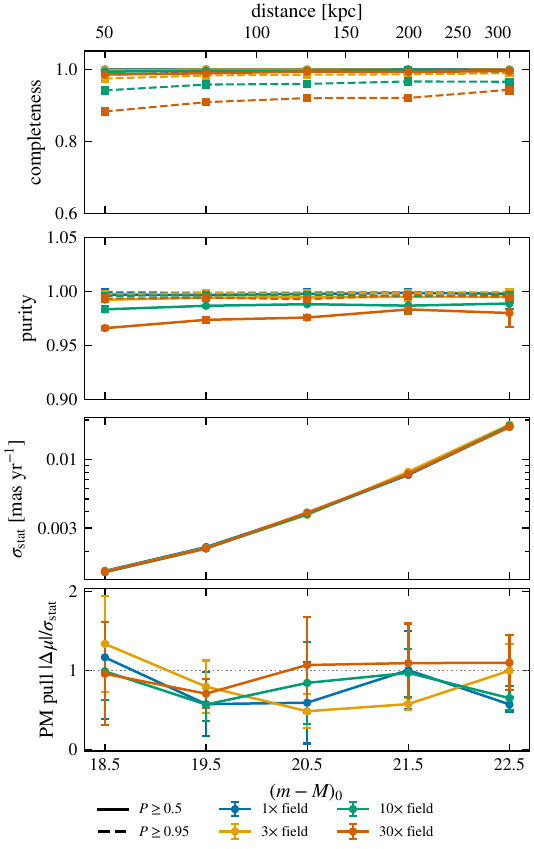}
\caption{Same as Fig.~\ref{fig:study} with the expected \Gaia{} DR4 uncertainties (PyGaia performance models for a 66-month baseline), everything else unchanged.}
\label{fig:study_dr4}
\end{center}
\end{figure}

Figure~\ref{fig:study_dr4} repeats the grid of Fig.~\ref{fig:study} with the uncertainties expected for \Gaia{} DR4, everything else unchanged; the results are discussed in Sect.~\ref{sec:validation}.

\section{A simplified test on a synthetic catalog}\label{apx:toy}

\begin{figure}
\begin{center}
\includegraphics[width=\columnwidth]{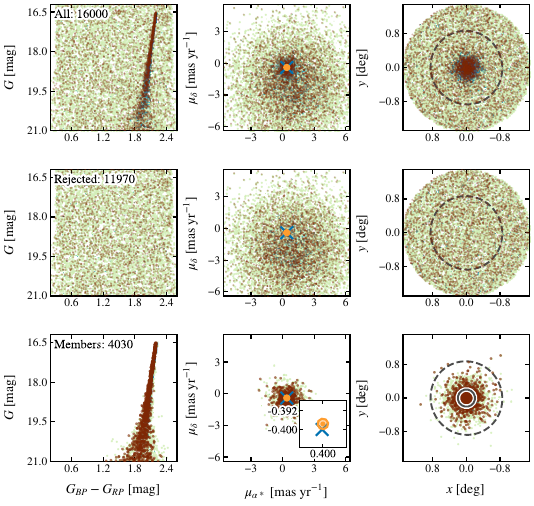}
\caption{Recovery of a synthetic system with the full model, including the $v_{\rm los}$ term, in the layout used for the real systems of Sect.~\ref{sec:application}. From top to bottom: all the sources of the cone, the sources rejected as non-members ($P < 0.5$), and the selected members; from left to right: CMD, PM plane and tangent-plane positions, as logarithmic density maps with the cells holding a single source drawn as points. The orange dot and filled ellipse are the recovered systemic PM and its $1\sigma$ statistical covariance, zoomed in the inset of the bottom-middle panel; the mock has no systematic term (no quasars are simulated), so the statistical ellipse is the total one. The blue cross is the input PM, the dashed ellipse the boundary of the object region, and the solid ellipse in the bottom panel the half-light ellipse of the fitted profile. The brick-red dots mark the stars with a line-of-sight velocity (30\% of the sample). This simplified sample (a straight RGB-like sequence, a Gaussian field) contains 4000 members and 12\,000 field stars in a cone of $1.5\degree$.}
\label{fig:synthetic}
\end{center}
\end{figure}

Figure~\ref{fig:synthetic} shows a first, simplified test of the model: 4000 members on a straight RGB-like sequence, with an exponential profile, immersed in a synthetic field of 12\,000 stars with broad PMs and parallaxes, in a cone of $1.5\degree$. A random 30\% of the stars, members and field alike, have a line-of-sight velocity with an error of $2\kms$ (members at $55\kms$ with a dispersion of $10\kms$, field stars spread over $90\kms$). Without the $v_{\rm los}$, the full model recovers 99.7\% of the members with a purity of 98.7\%; the 14 missed members and 40 of the 53 accepted field stars are fainter than $G \simeq 20$. The systemic PM, the scale radius of the profile and the expected number of members are recovered within one statistical error. With the $v_{\rm los}$, the stars that have one are selected with 99.8\% completeness and 99.8\% purity, the stars without one do not change, and the model returns $v_{\rm los,sys} = 55.3 \pm 0.3\kms$ with a dispersion of $9.9\kms$. Removing the photometric term lowers the completeness and the purity by about one percent, removing the spatial term lowers the purity by three percent, and the PM-only model reaches 93.9\% completeness and 91.9\% purity. This test is benevolent, since the sequence is bright and the field is Gaussian; the realistic mock catalogs of Sect.~\ref{sec:validation} are the actual validation.

\section{Diagnostic figures of NGC\,5466, Hercules and M67}\label{apx:examples}

\begin{figure}
\begin{center}
\includegraphics[width=\columnwidth]{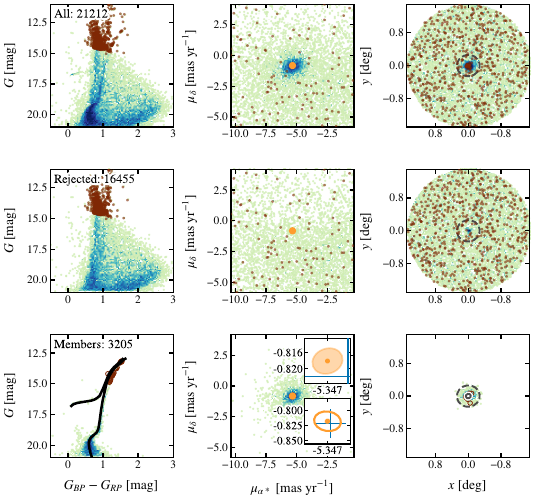}
\caption{Same as Fig.~\ref{fig:fornax} for the globular cluster NGC\,5466; the literature PM is that of \citet{Vasiliev2021}, whose errors include their systematic floor of $0.025\masyr$ and are therefore drawn as the thin (total) error bar only, comparable with the outer ellipse, $(0.022, 0.016)\masyr$ here, of the same size; the statistical ellipse, $(0.0037, 0.0032)\masyr$, is the small filled dot at its center, so that a single inset suffices. The isochrone has 12.5\,Gyr and ${\rm [Fe/H]} = -2.1$. The 1552 stars above the threshold that fail the quality screening, mostly in the crowded center, are not shown in the middle row.}
\label{fig:ngc5466}
\end{center}
\end{figure}

\begin{figure}
\begin{center}
\includegraphics[width=\columnwidth]{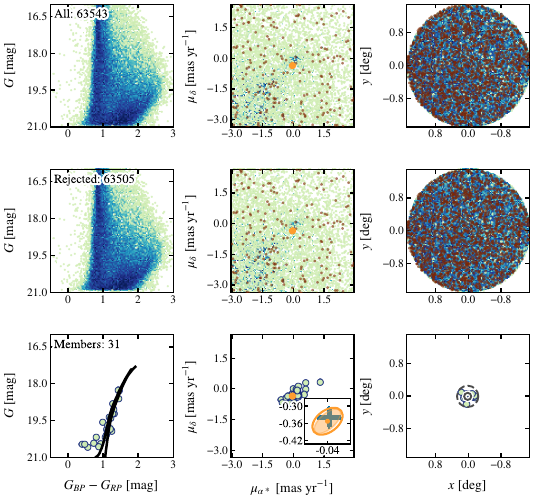}
\caption{Same as Fig.~\ref{fig:fornax} for the ultra-faint dwarf galaxy Hercules; the literature PM is that of \citet{Pace2022}, and the isochrone has 10\,Gyr and ${\rm [Fe/H]} = -1.6$. With only 31 members the density is too low for the color scale, so the members are drawn as individual points with a dark edge to make them visible.}
\label{fig:hercules}
\end{center}
\end{figure}

\begin{figure}
\begin{center}
\includegraphics[width=\columnwidth]{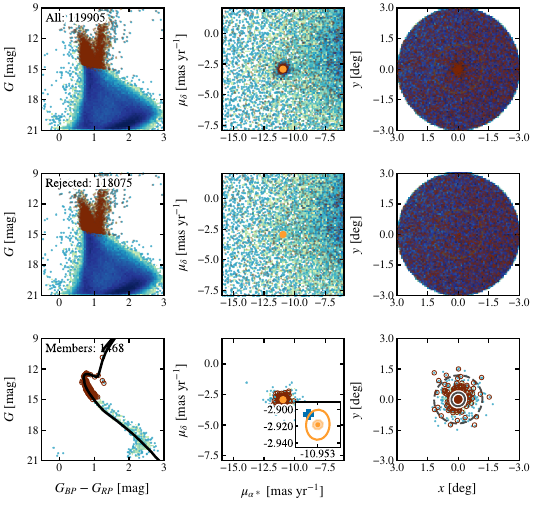}
\caption{Same as Fig.~\ref{fig:fornax} for the open cluster M67 (NGC\,2682); the literature PM is that of \citet[HR23]{HuntReffert2023}, with the error of the mean of their members as the only bar. The isochrone has 4\,Gyr and solar metallicity, drawn at the literature age, distance and reddening rather than fitted. The 362 stars above the threshold that fail the quality screening are mostly bright stars and binaries with a large RUWE.}
\label{fig:m67}
\end{center}
\end{figure}

Figures~\ref{fig:ngc5466} to \ref{fig:m67} show, for NGC\,5466, Hercules and M67, the same diagnostic panels as Fig.~\ref{fig:fornax} for Fornax: all the sources of the cone, the sources rejected as non-members and the selected members, in the CMD, the PM plane and the sky. The four runs are discussed in Sect.~\ref{sec:examples}.

\section{Systemic motions of the Local Group dwarf galaxies}\label{apx:lg}

\begin{table*}
\caption{Systemic PMs of the Local Group dwarf galaxies ($P \geq 0.9$, quasar systematics): the tier of the result (Sect.~\ref{sec:application}; 1, the position alone; 2, with the literature distance; 3, with the literature PM and the structural parameters of the LVDB; 4, a dedicated run), literature distance, number of members, detection significance, systemic PM with statistical and systematic errors, the correlation coefficient $\rho$ between the two components of the total (statistical plus systematic) covariance, and the literature PM with its quoted errors as compiled in the LVDB \citep{Pace2025}; the superscript letters give its source (legend below the table). The errors of \citet{Pace2022} are statistical; the other sources quote total errors. Systems marked with $\dagger$ are marginal detections ($2$--$3\sigma$), excluded from the statistics of Sect.~\ref{sec:localgroup}; those marked with $\ast$ are dedicated runs whose expected contamination lies between 10 and 20 percent.}
\label{tab:lg}
\centering
\scriptsize
\setlength{\tabcolsep}{3.5pt}
\renewcommand{\arraystretch}{0.9}
\begin{tabular}{l c r r r l l r c}
\toprule
Galaxy & Tier & $D$ & $N_{\rm mem}$ & $S$ & $\pmra$ & $\pmdec$ & $\rho$ & Literature $(\pmra, \pmdec)$ \\
       &  & [kpc] &  &  & [$\masyr$] & [$\masyr$] &  & [$\masyr$] \\
\midrule
Draco II & 1 & 20 & 28 & 7.6 & $+0.980 \pm 0.038 \pm 0.019$ & $+0.885 \pm 0.043 \pm 0.015$ & -0.00 & $(+1.043 \pm 0.029, +0.879 \pm 0.028)$$^{\rm a}$ \\
Segue 1 & 3 & 23 & 10 & 6.1 & $-2.049 \pm 0.053 \pm 0.020$ & $-3.434 \pm 0.041 \pm 0.025$ & -0.24 & $(-2.102 \pm 0.051, -3.375 \pm 0.045)$$^{\rm b}$ \\
Tucana III & 3 & 25 & 25 & 9.6 & $-0.031 \pm 0.035 \pm 0.017$ & $-1.543 \pm 0.057 \pm 0.017$ & +0.06 & $(-0.048 \pm 0.036, -1.638 \pm 0.039)$$^{\rm b}$ \\
Hydrus I & 1 & 28 & 86 & 13.0 & $+3.785 \pm 0.017 \pm 0.014$ & $-1.500 \pm 0.015 \pm 0.015$ & +0.19 & $(+3.781 \pm 0.016, -1.496 \pm 0.015)$$^{\rm b}$ \\
Carina III & 3 & 28 & 7 & 5.5 & $+3.085 \pm 0.042 \pm 0.020$ & $+1.364 \pm 0.072 \pm 0.018$ & +0.17 & $(+3.095 \pm 0.041, +1.395 \pm 0.045)$$^{\rm b}$ \\
Triangulum II & 3 & 30 & 6 & 3.3 & $+0.598 \pm 0.060 \pm 0.023$ & $+0.100 \pm 0.068 \pm 0.021$ & +0.19 & $(+0.575 \pm 0.060, +0.112 \pm 0.068)$$^{\rm b}$ \\
Cetus II & 3 & 30 & 2 & 3.1 & $+2.850 \pm 0.062 \pm 0.034$ & $+0.473 \pm 0.068 \pm 0.018$ & +0.23 & $(+2.844 \pm 0.060, +0.474 \pm 0.064)$$^{\rm b}$ \\
Reticulum II & 3 & 30 & 50 & 16.1 & $+2.356 \pm 0.032 \pm 0.016$ & $-1.354 \pm 0.027 \pm 0.019$ & -0.32 & $(+2.377 \pm 0.024, -1.379 \pm 0.026)$$^{\rm b}$ \\
Ursa Major II & 1 & 32 & 38 & 9.6 & $+1.745 \pm 0.021 \pm 0.024$ & $-1.895 \pm 0.024 \pm 0.014$ & +0.13 & $(+1.731 \pm 0.021, -1.906 \pm 0.025)$$^{\rm b}$ \\
Segue 2 & 1 & 35 & 15 & 6.6 & $+1.397 \pm 0.064 \pm 0.019$ & $-0.364 \pm 0.055 \pm 0.028$ & +0.14 & $(+1.446 \pm 0.059, -0.322 \pm 0.050)$$^{\rm b}$ \\
Carina II & 3 & 36 & 37 & 7.7 & $+1.893 \pm 0.023 \pm 0.017$ & $+0.151 \pm 0.022 \pm 0.015$ & -0.03 & $(+1.885 \pm 0.018, +0.133 \pm 0.019)$$^{\rm b}$ \\
Willman 1 & 3 & 38 & 9 & 6.5 & $+0.273 \pm 0.086 \pm 0.016$ & $-1.114 \pm 0.081 \pm 0.017$ & -0.11 & $(+0.299 \pm 0.052, -1.097 \pm 0.075)$$^{\rm c}$ \\
Boötes II & 1 & 42 & 22 & 7.8 & $-2.465 \pm 0.074 \pm 0.018$ & $-0.461 \pm 0.059 \pm 0.017$ & -0.14 & $(-2.426 \pm 0.079, -0.414 \pm 0.061)$$^{\rm b}$ \\
Coma Berenices & 1 & 44 & 37 & 9.7 & $+0.352 \pm 0.034 \pm 0.018$ & $-1.736 \pm 0.031 \pm 0.014$ & -0.02 & $(+0.423 \pm 0.026, -1.721 \pm 0.024)$$^{\rm b}$ \\
Boötes III$^\ast$ & 4 & 47 & 45 & 5.8 & $-1.124 \pm 0.017 \pm 0.028$ & $-0.857 \pm 0.013 \pm 0.010$ & -0.02 & $(-1.176 \pm 0.019, -0.890 \pm 0.015)$$^{\rm b}$ \\
Tucana IV$^\dagger$ & 3 & 48 & 9 & 2.5 & $+0.586 \pm 0.045 \pm 0.016$ & $-1.593 \pm 0.061 \pm 0.017$ & -0.01 & $(+0.534 \pm 0.052, -1.707 \pm 0.054)$$^{\rm b}$ \\
Grus II & 4 & 53 & 17 & 6.2 & $+0.389 \pm 0.030 \pm 0.014$ & $-1.465 \pm 0.034 \pm 0.014$ & +0.16 & $(+0.384 \pm 0.033, -1.484 \pm 0.040)$$^{\rm b}$ \\
Tucana II$^\ast$ & 4 & 58 & 26 & 4.4 & $+0.883 \pm 0.018 \pm 0.012$ & $-1.281 \pm 0.022 \pm 0.014$ & -0.14 & $(+0.911 \pm 0.025, -1.280 \pm 0.029)$$^{\rm b}$ \\
Boötes I & 1 & 66 & 184 & 23.7 & $-0.397 \pm 0.017 \pm 0.019$ & $-1.067 \pm 0.015 \pm 0.013$ & -0.04 & $(-0.385 \pm 0.017, -1.068 \pm 0.013)$$^{\rm b}$ \\
Sagittarius II & 1 & 70 & 53 & 18.1 & $-0.778 \pm 0.038 \pm 0.020$ & $-0.906 \pm 0.025 \pm 0.018$ & -0.02 & $(-0.769 \pm 0.035, -0.903 \pm 0.022)$$^{\rm b}$ \\
DELVE 2 & 3 & 71 & 7 & 5.0 & $+0.964 \pm 0.121 \pm 0.018$ & $-0.998 \pm 0.097 \pm 0.018$ & +0.19 & $(+0.920 \pm 0.115, -0.970 \pm 0.085)$$^{\rm d}$ \\
Draco & 1 & 76 & 1568 & 186.5 & $+0.043 \pm 0.006 \pm 0.015$ & $-0.189 \pm 0.007 \pm 0.014$ & +0.08 & $(+0.044 \pm 0.005, -0.188 \pm 0.006)$$^{\rm b}$ \\
Ursa Minor & 1 & 76 & 1915 & 192.6 & $-0.128 \pm 0.006 \pm 0.014$ & $+0.071 \pm 0.006 \pm 0.013$ & -0.13 & $(-0.120 \pm 0.005, +0.071 \pm 0.005)$$^{\rm b}$ \\
Eridanus IV & 3 & 77 & 31 & 7.1 & $+0.255 \pm 0.049 \pm 0.023$ & $-0.116 \pm 0.043 \pm 0.020$ & -0.02 & $(+0.250 \pm 0.060, -0.100 \pm 0.050)$$^{\rm e}$ \\
Horologium I & 1 & 79 & 17 & 8.2 & $+0.899 \pm 0.048 \pm 0.017$ & $-0.590 \pm 0.040 \pm 0.018$ & +0.12 & $(+0.847 \pm 0.035, -0.607 \pm 0.035)$$^{\rm b}$ \\
Phoenix II & 3 & 84 & 8 & 6.5 & $+0.501 \pm 0.052 \pm 0.017$ & $-1.198 \pm 0.063 \pm 0.018$ & -0.35 & $(+0.507 \pm 0.048, -1.199 \pm 0.058)$$^{\rm b}$ \\
Sextans & 1 & 86 & 1307 & 85.5 & $-0.406 \pm 0.009 \pm 0.016$ & $+0.017 \pm 0.009 \pm 0.016$ & -0.38 & $(-0.409 \pm 0.009, +0.037 \pm 0.009)$$^{\rm b}$ \\
Sculptor & 1 & 86 & 6604 & 1072.2 & $+0.101 \pm 0.003 \pm 0.019$ & $-0.154 \pm 0.002 \pm 0.015$ & -0.48 & $(+0.100 \pm 0.002, -0.158 \pm 0.002)$$^{\rm b}$ \\
Pegasus IV & 3 & 90 & 13 & 6.6 & $+0.334 \pm 0.086 \pm 0.020$ & $-0.222 \pm 0.091 \pm 0.018$ & +0.17 & $(+0.330 \pm 0.070, -0.210 \pm 0.080)$$^{\rm f}$ \\
Ursa Major I & 1 & 97 & 62 & 13.9 & $-0.395 \pm 0.044 \pm 0.013$ & $-0.740 \pm 0.039 \pm 0.013$ & -0.03 & $(-0.401 \pm 0.036, -0.613 \pm 0.041)$$^{\rm b}$ \\
Boötes V & 3 & 100 & 6 & 4.7 & $-0.213 \pm 0.059 \pm 0.017$ & $-0.275 \pm 0.084 \pm 0.018$ & +0.11 & $(-0.220 \pm 0.050, -0.280 \pm 0.070)$$^{\rm g}$ \\
Carina & 1 & 105 & 2044 & 108.2 & $+0.545 \pm 0.007 \pm 0.014$ & $+0.123 \pm 0.006 \pm 0.015$ & -0.06 & $(+0.532 \pm 0.007, +0.127 \pm 0.006)$$^{\rm b}$ \\
Aquarius II & 3 & 108 & 3 & 3.2 & $-0.079 \pm 0.420 \pm 0.027$ & $+0.346 \pm 0.345 \pm 0.019$ & +0.04 & $(-0.179 \pm 0.116, -0.466 \pm 0.096)$$^{\rm b}$ \\
Pictor I & 3 & 114 & 7 & 6.6 & $+0.158 \pm 0.093 \pm 0.020$ & $+0.054 \pm 0.122 \pm 0.022$ & -0.15 & $(+0.153 \pm 0.087, +0.096 \pm 0.116)$$^{\rm b}$ \\
Centaurus I & 3 & 116 & 12 & 7.8 & $-0.126 \pm 0.054 \pm 0.020$ & $-0.220 \pm 0.049 \pm 0.018$ & -0.04 & $(-0.140 \pm 0.050, -0.190 \pm 0.040)$$^{\rm h}$ \\
Crater II & 3 & 118 & 200 & 16.3 & $-0.071 \pm 0.020 \pm 0.016$ & $-0.110 \pm 0.013 \pm 0.011$ & +0.08 & $(-0.072 \pm 0.020, -0.112 \pm 0.013)$$^{\rm b}$ \\
Hercules & 1 & 132 & 31 & 10.4 & $-0.038 \pm 0.047 \pm 0.025$ & $-0.352 \pm 0.043 \pm 0.019$ & +0.43 & $(-0.035 \pm 0.042, -0.339 \pm 0.036)$$^{\rm b}$ \\
Antlia II & 1 & 132 & 243 & 10.7 & $-0.135 \pm 0.007 \pm 0.014$ & $+0.036 \pm 0.008 \pm 0.012$ & +0.33 & $(-0.093 \pm 0.008, +0.100 \pm 0.009)$$^{\rm b}$ \\
Hydra II & 3 & 134 & 13 & 7.5 & $-0.454 \pm 0.144 \pm 0.022$ & $-0.073 \pm 0.111 \pm 0.018$ & +0.03 & $(-0.394 \pm 0.140, +0.000 \pm 0.103)$$^{\rm b}$ \\
Fornax & 1 & 147 & 19982 & 2321.7 & $+0.382 \pm 0.001 \pm 0.010$ & $-0.359 \pm 0.002 \pm 0.011$ & -0.02 & $(+0.381 \pm 0.001, -0.359 \pm 0.002)$$^{\rm b}$ \\
Leo IV & 3 & 154 & 5 & 3.2 & $-0.150 \pm 0.141 \pm 0.022$ & $-0.295 \pm 0.108 \pm 0.021$ & -0.14 & $(-0.192 \pm 0.051, -0.069 \pm 0.052)$$^{\rm i}$ \\
Canes Venatici II & 1 & 160 & 14 & 12.7 & $-0.157 \pm 0.121 \pm 0.020$ & $-0.255 \pm 0.089 \pm 0.027$ & +0.32 & $(-0.124 \pm 0.116, -0.254 \pm 0.081)$$^{\rm b}$ \\
Leo V & 3 & 178 & 7 & 4.4 & $+0.077 \pm 0.240 \pm 0.021$ & $-0.328 \pm 0.173 \pm 0.021$ & -0.15 & $(+0.119 \pm 0.194, -0.118 \pm 0.170)$$^{\rm i}$ \\
Columba I & 3 & 183 & 5 & 3.9 & $+0.202 \pm 0.072 \pm 0.019$ & $-0.417 \pm 0.081 \pm 0.019$ & -0.10 & $(+0.169 \pm 0.072, -0.400 \pm 0.079)$$^{\rm b}$ \\
Boötes IV$^\dagger$ & 3 & 209 & 2 & 2.5 & $+0.210 \pm 0.197 \pm 0.018$ & $+0.074 \pm 0.258 \pm 0.018$ & +0.21 & $(+0.469 \pm 0.212, +0.489 \pm 0.256)$$^{\rm b}$ \\
Canes Venatici I & 1 & 218 & 142 & 38.5 & $-0.102 \pm 0.032 \pm 0.014$ & $-0.112 \pm 0.021 \pm 0.013$ & +0.15 & $(-0.096 \pm 0.030, -0.116 \pm 0.020)$$^{\rm b}$ \\
Leo II & 1 & 233 & 313 & 167.5 & $-0.107 \pm 0.030 \pm 0.022$ & $-0.140 \pm 0.028 \pm 0.022$ & -0.30 & $(-0.143 \pm 0.009, -0.130 \pm 0.009)$$^{\rm j}$ \\
Leo I & 1 & 254 & 1191 & 529.9 & $-0.046 \pm 0.016 \pm 0.019$ & $-0.116 \pm 0.011 \pm 0.032$ & -0.38 & $(-0.063 \pm 0.005, -0.111 \pm 0.004)$$^{\rm j}$ \\
Eridanus II & 1 & 366 & 17 & 10.1 & $+0.133 \pm 0.108 \pm 0.017$ & $+0.082 \pm 0.130 \pm 0.017$ & -0.15 & $(+0.125 \pm 0.101, +0.013 \pm 0.125)$$^{\rm b}$ \\
Phoenix & 1 & 415 & 207 & 136.9 & $+0.105 \pm 0.032 \pm 0.016$ & $-0.076 \pm 0.038 \pm 0.017$ & -0.11 & $(+0.142 \pm 0.028, -0.040 \pm 0.032)$$^{\rm j}$ \\
Leo T & 3 & 417 & 6 & 5.2 & $+0.051 \pm 0.445 \pm 0.025$ & $-0.110 \pm 0.262 \pm 0.020$ & -0.37 & $(-0.026 \pm 0.015, -0.129 \pm 0.015)$$^{\rm k}$ \\
NGC 6822 & 1 & 490 & 1382 & 135.0 & $-0.048 \pm 0.009 \pm 0.020$ & $-0.072 \pm 0.008 \pm 0.019$ & +0.12 & $(-0.060 \pm 0.010, -0.070 \pm 0.010)$$^{\rm d}$ \\
NGC 185$^\dagger$ & 3 & 617 & 2 & 2.1 & $-0.035 \pm 0.125 \pm 0.020$ & $-0.012 \pm 0.097 \pm 0.021$ & -0.49 & $(+0.024 \pm 0.014, +0.006 \pm 0.015)$$^{\rm l}$ \\
IC 1613 & 1 & 720 & 584 & 232.0 & $+0.043 \pm 0.020 \pm 0.018$ & $-0.013 \pm 0.015 \pm 0.018$ & +0.20 & $(+0.040 \pm 0.020, +0.010 \pm 0.010)$$^{\rm d}$ \\
IC 10 & 2 & 794 & 29 & 11.1 & $+0.081 \pm 0.040 \pm 0.034$ & $-0.097 \pm 0.043 \pm 0.034$ & -0.25 & $(-0.002 \pm 0.008, +0.020 \pm 0.008)$$^{\rm m}$ \\
Leo A & 1 & 800 & 46 & 38.5 & $-0.183 \pm 0.091 \pm 0.018$ & $-0.082 \pm 0.088 \pm 0.018$ & -0.25 & $(+0.002 \pm 0.097, -0.084 \pm 0.090)$$^{\rm j}$ \\
M 32 & 1 & 805 & 6754 & 182.6 & $+0.029 \pm 0.004 \pm 0.011$ & $-0.041 \pm 0.003 \pm 0.011$ & +0.01 & -- \\
NGC 205 & 1 & 824 & 6176 & 141.7 & $+0.022 \pm 0.003 \pm 0.012$ & $-0.045 \pm 0.003 \pm 0.010$ & -0.06 & -- \\
\bottomrule
\end{tabular}\\[3pt]
\parbox{0.95\textwidth}{\footnotesize Sources of the literature PMs: (a) \citet{Warfield2026}; (b) \citet{Pace2022}; (c) \citet{Chiu2026}; (d) \citet{Battaglia2022}; (e) \citet{Cerny2021}; (f) \citet{Cerny2023a}; (g) \citet{Cerny2023}; (h) \citet{Heiger2024}; (i) \citet{Julio2024}; (j) \citet{Bennet2024}; (k) \citet{Bennet2025}; (l) \citet{Sohn2020}; (m) \citet{Brunthaler2007}.}
\end{table*}

Table~\ref{tab:lg} lists the systemic PMs of the dwarf galaxies recovered in Sect.~\ref{sec:localgroup}, together with the three marginal systems.

\section{Systemic motions of the globular clusters}\label{apx:gcs}

\begin{table*}
\caption{Systemic PMs of the Galactic globular clusters ($P \geq 0.9$, quasar systematics): the tier of the result (Sect.~\ref{sec:application}; 1, the position alone; 2, with the distance of \citet{Harris1996}; 3, with the literature PM and the half-light radius of the catalog; 4, a dedicated run), number of members, detection significance, systemic PM with statistical and systematic errors, the correlation coefficient $\rho$ between the two components of the total (statistical plus systematic) covariance, the PM of \citet{Vasiliev2021} with its quoted error, which includes a systematic floor of $0.025\masyr$, the systemic line-of-sight velocity from the \Gaia{} RVS velocities of the members (total error; the number of members with a $v_{\rm los}$ in parentheses), and the $v_{\rm los,sys}$ of the compilation of \citet{Baumgardt2019}. First half, in order of right ascension. Flags on the name as in Table~\ref{tab:lg}.}
\label{tab:gcs_a}
\centering
\footnotesize
\setlength{\tabcolsep}{3pt}
\renewcommand{\arraystretch}{0.95}
\resizebox{\textwidth}{!}{%
\begin{tabular}{l c r r l l r c l l}
\toprule
Cluster & Tier & $N_{\rm mem}$ & $S$ & $\pmra$ & $\pmdec$ & $\rho$ & VB21 $(\pmra, \pmdec)$ & $v_{\rm los,sys}$ ($N_v$) & $v_{\rm los,sys}^{\rm lit}$ \\
        &  &  &  & [$\masyr$] & [$\masyr$] &  & [$\masyr$] & [$\kms$] & [$\kms$] \\
\midrule
NGC 104 & 1 & 55807 & 2153 & $+5.258 \pm 0.002 \pm 0.014$ & $-2.555 \pm 0.002 \pm 0.014$ & +0.03 & $(+5.252 \pm 0.021, -2.551 \pm 0.021)$ & $-17.1 \pm 0.4$ (908) & $-17.4 \pm 0.2$ \\
NGC 288 & 1 & 5869 & 1656 & $+4.156 \pm 0.003 \pm 0.027$ & $-5.707 \pm 0.003 \pm 0.021$ & +0.49 & $(+4.164 \pm 0.024, -5.705 \pm 0.024)$ & $-44.7 \pm 0.6$ (51) & $-44.5 \pm 0.1$ \\
NGC 362 & 1 & 6634 & 722 & $+6.689 \pm 0.004 \pm 0.016$ & $-2.538 \pm 0.003 \pm 0.018$ & -0.01 & $(+6.694 \pm 0.025, -2.535 \pm 0.024)$ & $+223.0 \pm 0.8$ (39) & $+223.1 \pm 0.3$ \\
NGC 1261 & 1 & 2061 & 566 & $+1.603 \pm 0.004 \pm 0.017$ & $-2.062 \pm 0.005 \pm 0.018$ & -0.17 & $(+1.596 \pm 0.025, -2.064 \pm 0.025)$ & $+71.8 \pm 0.9$ (17) & $+71.3 \pm 0.2$ \\
Pal 1 & 1 & 81 & 36 & $-0.256 \pm 0.024 \pm 0.020$ & $+0.002 \pm 0.027 \pm 0.020$ & +0.12 & $(-0.252 \pm 0.034, +0.007 \pm 0.037)$ & -- & $-75.7 \pm 0.3$ \\
AM 1 & 1 & 38 & 34 & $+0.347 \pm 0.071 \pm 0.019$ & $-0.181 \pm 0.088 \pm 0.019$ & -0.18 & $(+0.291 \pm 0.107, -0.177 \pm 0.086)$ & -- & $+118.0 \pm 14.1$ \\
Pal 2 & 1 & 157 & 43 & $+1.047 \pm 0.021 \pm 0.026$ & $-1.528 \pm 0.017 \pm 0.022$ & +0.00 & $(+1.045 \pm 0.034, -1.522 \pm 0.031)$ & -- & $-136.0 \pm 1.6$ \\
NGC 1851 & 1 & 4594 & 775 & $+2.138 \pm 0.003 \pm 0.017$ & $-0.656 \pm 0.004 \pm 0.021$ & -0.22 & $(+2.145 \pm 0.024, -0.650 \pm 0.024)$ & $+320.7 \pm 1.0$ (32) & $+321.4 \pm 1.6$ \\
NGC 1904 & 1 & 2683 & 637 & $+2.468 \pm 0.004 \pm 0.016$ & $-1.598 \pm 0.005 \pm 0.024$ & +0.16 & $(+2.469 \pm 0.025, -1.594 \pm 0.025)$ & $+208.2 \pm 0.7$ (26) & $+205.8 \pm 0.2$ \\
NGC 2298 & 1 & 1556 & 460 & $+3.313 \pm 0.005 \pm 0.029$ & $-2.171 \pm 0.005 \pm 0.018$ & +0.22 & $(+3.320 \pm 0.025, -2.175 \pm 0.026)$ & $+147.8 \pm 1.0$ (18) & $+147.2 \pm 0.6$ \\
NGC 2419 & 1 & 752 & 181 & $-0.001 \pm 0.010 \pm 0.017$ & $-0.523 \pm 0.007 \pm 0.020$ & +0.01 & $(+0.007 \pm 0.028, -0.523 \pm 0.026)$ & -- & $-21.1 \pm 0.3$ \\
Pyxis Cluster & 1 & 128 & 39 & $+1.020 \pm 0.016 \pm 0.019$ & $+0.152 \pm 0.019 \pm 0.023$ & +0.09 & $(+1.030 \pm 0.032, +0.138 \pm 0.035)$ & -- & -- \\
NGC 2808 & 1 & 7962 & 591 & $+0.993 \pm 0.004 \pm 0.018$ & $+0.272 \pm 0.004 \pm 0.021$ & +0.14 & $(+0.994 \pm 0.024, +0.273 \pm 0.024)$ & $+103.6 \pm 1.2$ (77) & $+103.6 \pm 0.3$ \\
E 3 & 1 & 447 & 97 & $-2.723 \pm 0.010 \pm 0.019$ & $+7.093 \pm 0.009 \pm 0.019$ & +0.05 & $(-2.727 \pm 0.027, +7.083 \pm 0.027)$ & -- & $+11.7 \pm 0.4$ \\
Pal 3 & 1 & 47 & 38 & $+0.092 \pm 0.057 \pm 0.027$ & $-0.177 \pm 0.068 \pm 0.021$ & -0.44 & $(+0.086 \pm 0.060, -0.148 \pm 0.071)$ & -- & $+94.0 \pm 0.8$ \\
NGC 3201 & 1 & 19875 & 4552 & $+8.343 \pm 0.002 \pm 0.016$ & $-1.963 \pm 0.002 \pm 0.016$ & -0.11 & $(+8.348 \pm 0.022, -1.958 \pm 0.022)$ & $+495.2 \pm 0.4$ (148) & $+495.4 \pm 0.1$ \\
Pal 4 & 1 & 80 & 68 & $-0.196 \pm 0.032 \pm 0.022$ & $-0.472 \pm 0.031 \pm 0.024$ & -0.30 & $(-0.188 \pm 0.042, -0.476 \pm 0.041)$ & -- & $+72.4 \pm 0.2$ \\
Crater Cluster$^\dagger$ & 3 & 4 & 3 & $-0.003 \pm 0.118 \pm 0.022$ & $-0.209 \pm 0.093 \pm 0.020$ & -0.21 & $(-0.059 \pm 0.125, -0.116 \pm 0.116)$ & -- & -- \\
NGC 4147 & 1 & 523 & 248 & $-1.707 \pm 0.010 \pm 0.020$ & $-2.087 \pm 0.009 \pm 0.021$ & +0.17 & $(-1.707 \pm 0.027, -2.090 \pm 0.027)$ & -- & $+179.3 \pm 0.3$ \\
NGC 4372 & 1 & 9815 & 397 & $-6.413 \pm 0.003 \pm 0.018$ & $+3.291 \pm 0.003 \pm 0.020$ & +0.02 & $(-6.409 \pm 0.024, +3.297 \pm 0.024)$ & $+75.5 \pm 0.6$ (146) & $+75.6 \pm 0.3$ \\
Rup 106 & 1 & 521 & 120 & $-1.264 \pm 0.007 \pm 0.021$ & $+0.399 \pm 0.006 \pm 0.018$ & -0.00 & $(-1.254 \pm 0.026, +0.401 \pm 0.026)$ & $-36.5 \pm 1.0$ (5) & $-38.4 \pm 0.3$ \\
NGC 4590 & 1 & 4424 & 896 & $-2.734 \pm 0.004 \pm 0.019$ & $+1.773 \pm 0.003 \pm 0.016$ & -0.01 & $(-2.739 \pm 0.024, +1.779 \pm 0.024)$ & $-92.8 \pm 0.8$ (42) & $-93.1 \pm 0.2$ \\
NGC 4833 & 1 & 3033 & 133 & $-8.388 \pm 0.004 \pm 0.020$ & $-0.973 \pm 0.004 \pm 0.019$ & +0.05 & $(-8.377 \pm 0.025, -0.963 \pm 0.025)$ & $+199.9 \pm 0.7$ (73) & $+202.0 \pm 0.4$ \\
NGC 5024 & 1 & 4915 & 657 & $-0.143 \pm 0.004 \pm 0.031$ & $-1.330 \pm 0.004 \pm 0.021$ & +0.51 & $(-0.133 \pm 0.024, -1.331 \pm 0.024)$ & $-61.7 \pm 0.9$ (35) & $-63.4 \pm 0.2$ \\
NGC 5053 & 1 & 1913 & 336 & $-0.329 \pm 0.006 \pm 0.033$ & $-1.213 \pm 0.006 \pm 0.021$ & +0.44 & $(-0.329 \pm 0.025, -1.213 \pm 0.025)$ & $+44.6 \pm 1.0$ (11) & $+42.8 \pm 0.2$ \\
NGC 5139 & 1 & 74343 & 2680 & $-3.254 \pm 0.002 \pm 0.014$ & $-6.764 \pm 0.002 \pm 0.021$ & -0.06 & $(-3.250 \pm 0.022, -6.746 \pm 0.022)$ & $+235.0 \pm 0.5$ (836) & $+232.8 \pm 0.2$ \\
NGC 5272 & 1 & 13481 & 2364 & $-0.152 \pm 0.003 \pm 0.034$ & $-2.675 \pm 0.002 \pm 0.016$ & -0.47 & $(-0.152 \pm 0.023, -2.670 \pm 0.022)$ & $-146.0 \pm 0.6$ (87) & $-147.2 \pm 0.3$ \\
NGC 5286 & 1 & 2353 & 257 & $+0.191 \pm 0.006 \pm 0.021$ & $-0.165 \pm 0.005 \pm 0.019$ & +0.06 & $(+0.198 \pm 0.025, -0.153 \pm 0.025)$ & $+61.2 \pm 1.1$ (37) & $+62.4 \pm 0.4$ \\
AM 4 & 3 & 6 & 5 & $-0.322 \pm 0.373 \pm 0.025$ & $-2.582 \pm 0.423 \pm 0.020$ & -0.46 & $(-0.291 \pm 0.445, -2.512 \pm 0.344)$ & -- & $+151.2 \pm 2.9$ \\
NGC 5466 & 1 & 3205 & 1051 & $-5.347 \pm 0.004 \pm 0.022$ & $-0.818 \pm 0.003 \pm 0.016$ & -0.10 & $(-5.342 \pm 0.025, -0.822 \pm 0.024)$ & $+108.1 \pm 0.6$ (30) & $+106.8 \pm 0.2$ \\
NGC 5634 & 1 & 520 & 203 & $-1.690 \pm 0.009 \pm 0.019$ & $-1.472 \pm 0.007 \pm 0.023$ & -0.03 & $(-1.692 \pm 0.027, -1.478 \pm 0.026)$ & $-18.6 \pm 2.4$ (5) & $-16.1 \pm 0.6$ \\
NGC 5694 & 1 & 286 & 110 & $-0.466 \pm 0.013 \pm 0.023$ & $-1.095 \pm 0.012 \pm 0.020$ & -0.19 & $(-0.464 \pm 0.029, -1.105 \pm 0.029)$ & -- & $-139.6 \pm 0.5$ \\
IC 4499 & 1 & 1323 & 308 & $+0.466 \pm 0.005 \pm 0.019$ & $-0.483 \pm 0.006 \pm 0.019$ & +0.05 & $(+0.466 \pm 0.025, -0.489 \pm 0.025)$ & $+38.0 \pm 1.1$ (12) & $+38.4 \pm 0.3$ \\
NGC 5824 & 1 & 992 & 110 & $-1.197 \pm 0.007 \pm 0.022$ & $-2.232 \pm 0.006 \pm 0.019$ & -0.13 & $(-1.189 \pm 0.026, -2.234 \pm 0.026)$ & $-16.2 \pm 8.6$ (5) & $-25.2 \pm 0.5$ \\
Pal 5 & 1 & 444 & 93 & $-2.727 \pm 0.012 \pm 0.025$ & $-2.654 \pm 0.010 \pm 0.020$ & +0.33 & $(-2.730 \pm 0.028, -2.654 \pm 0.027)$ & -- & $-58.6 \pm 0.1$ \\
NGC 5897 & 1 & 3524 & 468 & $-5.415 \pm 0.004 \pm 0.020$ & $-3.391 \pm 0.004 \pm 0.025$ & -0.19 & $(-5.422 \pm 0.025, -3.393 \pm 0.025)$ & $+101.8 \pm 1.0$ (38) & $+101.3 \pm 0.2$ \\
NGC 5904 & 1 & 13313 & 2591 & $+4.072 \pm 0.003 \pm 0.021$ & $-9.866 \pm 0.003 \pm 0.018$ & +0.32 & $(+4.086 \pm 0.023, -9.870 \pm 0.023)$ & $+52.8 \pm 0.8$ (105) & $+53.5 \pm 0.2$ \\
NGC 5927 & 3 & 303 & 22 & $-5.054 \pm 0.006 \pm 0.022$ & $-3.197 \pm 0.007 \pm 0.022$ & -0.00 & $(-5.056 \pm 0.025, -3.217 \pm 0.025)$ & $-104.8 \pm 0.7$ (46) & $-104.1 \pm 0.3$ \\
NGC 5986 & 1 & 1608 & 99 & $-4.186 \pm 0.006 \pm 0.022$ & $-4.562 \pm 0.006 \pm 0.022$ & +0.12 & $(-4.192 \pm 0.026, -4.568 \pm 0.026)$ & $+101.2 \pm 1.1$ (35) & $+101.2 \pm 0.4$ \\
Pal 14 & 1 & 99 & 30 & $-0.491 \pm 0.025 \pm 0.025$ & $-0.448 \pm 0.024 \pm 0.025$ & -0.22 & $(-0.463 \pm 0.038, -0.413 \pm 0.038)$ & -- & $+72.3 \pm 0.1$ \\
NGC 6093 & 1 & 1887 & 130 & $-2.939 \pm 0.008 \pm 0.021$ & $-5.585 \pm 0.007 \pm 0.021$ & +0.04 & $(-2.934 \pm 0.027, -5.578 \pm 0.026)$ & $+13.6 \pm 1.5$ (28) & $+10.9 \pm 0.4$ \\
NGC 6121 & 1 & 16722 & 2913 & $-12.516 \pm 0.004 \pm 0.020$ & $-19.011 \pm 0.004 \pm 0.020$ & +0.03 & $(-12.514 \pm 0.023, -19.022 \pm 0.023)$ & $+71.6 \pm 0.4$ (221) & $+71.2 \pm 0.1$ \\
NGC 6101 & 1 & 3943 & 679 & $+1.760 \pm 0.003 \pm 0.018$ & $-0.262 \pm 0.003 \pm 0.017$ & +0.12 & $(+1.756 \pm 0.024, -0.258 \pm 0.025)$ & $+365.4 \pm 0.9$ (35) & $+366.3 \pm 0.3$ \\
NGC 6144 & 1 & 1407 & 107 & $-1.747 \pm 0.007 \pm 0.022$ & $-2.610 \pm 0.006 \pm 0.022$ & -0.02 & $(-1.744 \pm 0.026, -2.607 \pm 0.026)$ & $+195.2 \pm 0.6$ (32) & $+194.8 \pm 0.6$ \\
NGC 6171 & 1 & 4008 & 357 & $-1.940 \pm 0.005 \pm 0.021$ & $-5.978 \pm 0.004 \pm 0.020$ & -0.02 & $(-1.939 \pm 0.025, -5.979 \pm 0.025)$ & $-33.6 \pm 0.6$ (60) & $-34.7 \pm 0.2$ \\
NGC 6205 & 1 & 13993 & 1352 & $-3.139 \pm 0.002 \pm 0.014$ & $-2.570 \pm 0.003 \pm 0.016$ & -0.07 & $(-3.149 \pm 0.023, -2.574 \pm 0.023)$ & $-246.7 \pm 0.7$ (102) & $-244.9 \pm 0.3$ \\
NGC 6229 & 1 & 419 & 165 & $-1.170 \pm 0.008 \pm 0.021$ & $-0.461 \pm 0.011 \pm 0.023$ & +0.36 & $(-1.171 \pm 0.026, -0.467 \pm 0.027)$ & $-138.0 \pm 2.0$ (8) & $-137.9 \pm 0.7$ \\
\bottomrule
\end{tabular}}
\end{table*}

\begin{figure*}[p]
\setcounter{apxnext}{\value{section}}\stepcounter{apxnext}\setcounter{figure}{0}\renewcommand{\thefigure}{\Alph{apxnext}.\arabic{figure}}
\begin{center}
\includegraphics[width=0.75\textwidth]{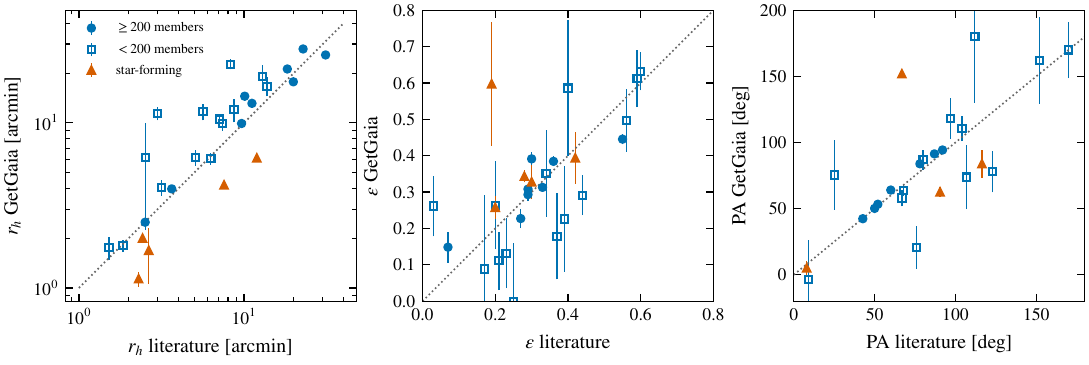}
\caption{Structural parameters of the 28 recovered dwarf galaxies whose spatial profile was fitted freely, compared with the photometric values compiled in the LVDB \citep{Pace2025}: half-light radius (left), ellipticity (middle) and position angle (right, for the systems with a literature ellipticity above 0.1; the smallest angle between the two orientations is plotted). The error bars are bootstrap uncertainties over the members (100 resamplings). Filled circles: systems with at least 200 members; open squares: fewer; triangles: star-forming dwarfs, whose \Gaia{} members are young stars. The values are listed in Table~\ref{tab:structure}.}
\label{fig:structure}
\end{center}
\end{figure*}

\begin{table*}
\caption{Same as Table~\ref{tab:gcs_a}, second half.}
\label{tab:gcs_b}
\centering
\footnotesize
\setlength{\tabcolsep}{3pt}
\renewcommand{\arraystretch}{0.95}
\resizebox{\textwidth}{!}{%
\begin{tabular}{l c r r l l r c l l}
\toprule
Cluster & Tier & $N_{\rm mem}$ & $S$ & $\pmra$ & $\pmdec$ & $\rho$ & VB21 $(\pmra, \pmdec)$ & $v_{\rm los,sys}$ ($N_v$) & $v_{\rm los,sys}^{\rm lit}$ \\
        &  &  &  & [$\masyr$] & [$\masyr$] &  & [$\masyr$] & [$\kms$] & [$\kms$] \\
\midrule
NGC 6218 & 1 & 8118 & 697 & $-0.200 \pm 0.003 \pm 0.020$ & $-6.804 \pm 0.003 \pm 0.038$ & -0.32 & $(-0.191 \pm 0.024, -6.802 \pm 0.024)$ & $-40.7 \pm 0.7$ (71) & $-41.7 \pm 0.1$ \\
NGC 6235 & 1 & 580 & 61 & $-3.937 \pm 0.009 \pm 0.022$ & $-7.601 \pm 0.007 \pm 0.022$ & +0.03 & $(-3.931 \pm 0.027, -7.587 \pm 0.027)$ & $+130.8 \pm 0.9$ (19) & $+126.7 \pm 0.3$ \\
NGC 6254 & 1 & 10117 & 604 & $-4.748 \pm 0.004 \pm 0.023$ & $-6.606 \pm 0.003 \pm 0.018$ & -0.09 & $(-4.758 \pm 0.024, -6.597 \pm 0.024)$ & $+76.0 \pm 0.6$ (112) & $+74.2 \pm 0.2$ \\
Pal 15 & 1 & 112 & 36 & $-0.593 \pm 0.027 \pm 0.021$ & $-0.881 \pm 0.022 \pm 0.031$ & -0.15 & $(-0.592 \pm 0.037, -0.901 \pm 0.034)$ & -- & $+72.3 \pm 1.7$ \\
NGC 6266 & 1 & 1085 & 38 & $-4.971 \pm 0.007 \pm 0.021$ & $-2.954 \pm 0.006 \pm 0.023$ & -0.02 & $(-4.978 \pm 0.026, -2.947 \pm 0.026)$ & $-72.0 \pm 0.8$ (139) & $-74.0 \pm 0.7$ \\
NGC 6273 & 1 & 2026 & 113 & $-3.252 \pm 0.007 \pm 0.021$ & $+1.672 \pm 0.006 \pm 0.021$ & -0.03 & $(-3.249 \pm 0.026, +1.660 \pm 0.025)$ & $+147.0 \pm 1.0$ (61) & $+145.5 \pm 0.6$ \\
NGC 6284 & 3 & 104 & 21 & $-3.205 \pm 0.010 \pm 0.022$ & $-2.011 \pm 0.010 \pm 0.022$ & +0.00 & $(-3.200 \pm 0.029, -2.002 \pm 0.028)$ & $+28.3 \pm 1.6$ (10) & $+28.6 \pm 0.7$ \\
NGC 6287 & 1 & 497 & 58 & $-5.000 \pm 0.009 \pm 0.022$ & $-1.876 \pm 0.008 \pm 0.020$ & +0.02 & $(-5.010 \pm 0.029, -1.883 \pm 0.028)$ & $-292.3 \pm 1.2$ (23) & $-294.7 \pm 1.6$ \\
NGC 6341 & 1 & 6728 & 1019 & $-4.936 \pm 0.004 \pm 0.016$ & $-0.635 \pm 0.004 \pm 0.015$ & -0.00 & $(-4.935 \pm 0.024, -0.625 \pm 0.024)$ & $-120.3 \pm 0.8$ (55) & $-120.5 \pm 0.3$ \\
NGC 6333 & 1 & 704 & 50 & $-2.171 \pm 0.007 \pm 0.022$ & $-3.218 \pm 0.005 \pm 0.021$ & +0.03 & $(-2.180 \pm 0.026, -3.222 \pm 0.026)$ & $+313.8 \pm 1.3$ (33) & $+310.8 \pm 2.1$ \\
NGC 6352 & 1 & 1809 & 71 & $-2.160 \pm 0.005 \pm 0.021$ & $-4.436 \pm 0.004 \pm 0.021$ & -0.00 & $(-2.158 \pm 0.025, -4.447 \pm 0.025)$ & $-124.6 \pm 0.6$ (81) & $-125.6 \pm 1.0$ \\
IC 1257 & 1 & 39 & 18 & $-1.025 \pm 0.027 \pm 0.023$ & $-1.493 \pm 0.021 \pm 0.025$ & +0.11 & $(-1.007 \pm 0.040, -1.492 \pm 0.032)$ & -- & $-138.0 \pm 2.0$ \\
NGC 6366 & 1 & 5676 & 321 & $-0.334 \pm 0.004 \pm 0.021$ & $-5.165 \pm 0.003 \pm 0.021$ & +0.01 & $(-0.332 \pm 0.025, -5.160 \pm 0.024)$ & $-121.3 \pm 0.4$ (149) & $-120.7 \pm 0.2$ \\
NGC 6362 & 1 & 7175 & 643 & $-5.508 \pm 0.002 \pm 0.019$ & $-4.768 \pm 0.002 \pm 0.016$ & -0.06 & $(-5.506 \pm 0.024, -4.763 \pm 0.024)$ & $-13.3 \pm 0.6$ (62) & $-14.6 \pm 0.2$ \\
NGC 6388 & 1 & 705 & 51 & $-1.318 \pm 0.007 \pm 0.022$ & $-2.704 \pm 0.006 \pm 0.022$ & -0.04 & $(-1.316 \pm 0.026, -2.709 \pm 0.026)$ & $+81.6 \pm 1.1$ (60) & $+83.1 \pm 0.5$ \\
NGC 6402 & 1 & 2503 & 92 & $-3.587 \pm 0.006 \pm 0.020$ & $-5.050 \pm 0.005 \pm 0.020$ & +0.02 & $(-3.590 \pm 0.025, -5.059 \pm 0.025)$ & $-60.0 \pm 1.1$ (67) & $-60.7 \pm 0.5$ \\
NGC 6397 & 1 & 20970 & 2967 & $+3.255 \pm 0.003 \pm 0.022$ & $-17.656 \pm 0.003 \pm 0.018$ & -0.29 & $(+3.260 \pm 0.023, -17.664 \pm 0.022)$ & $+19.1 \pm 0.5$ (182) & $+18.5 \pm 0.1$ \\
NGC 6426 & 1 & 258 & 51 & $-1.823 \pm 0.008 \pm 0.021$ & $-3.000 \pm 0.007 \pm 0.019$ & +0.04 & $(-1.828 \pm 0.026, -2.999 \pm 0.026)$ & $-209.1 \pm 2.3$ (6) & $-210.5 \pm 0.5$ \\
NGC 6496 & 1 & 1019 & 87 & $-3.064 \pm 0.005 \pm 0.023$ & $-9.266 \pm 0.005 \pm 0.021$ & +0.10 & $(-3.060 \pm 0.027, -9.271 \pm 0.026)$ & $-133.5 \pm 0.7$ (24) & $-134.7 \pm 0.3$ \\
NGC 6535 & 3 & 622 & 77 & $-4.215 \pm 0.008 \pm 0.022$ & $-2.941 \pm 0.007 \pm 0.023$ & +0.03 & $(-4.214 \pm 0.027, -2.939 \pm 0.026)$ & $-218.3 \pm 2.4$ (6) & $-214.8 \pm 0.5$ \\
NGC 6541 & 1 & 3504 & 163 & $+0.288 \pm 0.005 \pm 0.020$ & $-8.844 \pm 0.004 \pm 0.022$ & +0.09 & $(+0.287 \pm 0.025, -8.847 \pm 0.025)$ & $-166.8 \pm 0.7$ (67) & $-164.0 \pm 0.5$ \\
ESO 280-SC06 & 1 & 38 & 11 & $-0.711 \pm 0.022 \pm 0.021$ & $-2.763 \pm 0.017 \pm 0.022$ & +0.13 & $(-0.688 \pm 0.039, -2.777 \pm 0.033)$ & -- & $+93.2 \pm 0.3$ \\
NGC 6584 & 1 & 1214 & 124 & $-0.089 \pm 0.005 \pm 0.020$ & $-7.204 \pm 0.005 \pm 0.018$ & -0.09 & $(-0.090 \pm 0.026, -7.202 \pm 0.025)$ & $+260.7 \pm 1.6$ (18) & $+260.6 \pm 1.6$ \\
NGC 6652 & 1 & 399 & 49 & $-5.492 \pm 0.008 \pm 0.023$ & $-4.258 \pm 0.007 \pm 0.021$ & -0.04 & $(-5.484 \pm 0.027, -4.274 \pm 0.027)$ & $-96.8 \pm 1.4$ (11) & $-95.4 \pm 0.9$ \\
NGC 6656 & 1 & 10463 & 1791 & $+9.848 \pm 0.004 \pm 0.020$ & $-5.616 \pm 0.004 \pm 0.020$ & -0.00 & $(+9.851 \pm 0.023, -5.617 \pm 0.023)$ & $-147.5 \pm 0.5$ (328) & $-148.7 \pm 0.8$ \\
NGC 6681 & 1 & 896 & 76 & $+1.440 \pm 0.006 \pm 0.021$ & $-4.741 \pm 0.005 \pm 0.020$ & +0.00 & $(+1.431 \pm 0.027, -4.744 \pm 0.026)$ & $+216.0 \pm 1.0$ (16) & $+216.6 \pm 0.8$ \\
NGC 6715$^\ast$ & 4 & 128 & 18 & $-2.675 \pm 0.007 \pm 0.023$ & $-1.375 \pm 0.006 \pm 0.020$ & +0.09 & $(-2.679 \pm 0.025, -1.387 \pm 0.025)$ & $+139.3 \pm 2.3$ (14) & $+143.1 \pm 0.4$ \\
NGC 6717 & 1 & 225 & 17 & $-3.148 \pm 0.010 \pm 0.023$ & $-5.020 \pm 0.008 \pm 0.023$ & +0.13 & $(-3.125 \pm 0.027, -5.008 \pm 0.027)$ & $+28.6 \pm 1.3$ (9) & $+30.2 \pm 0.9$ \\
NGC 6723 & 1 & 2831 & 195 & $+1.031 \pm 0.005 \pm 0.019$ & $-2.419 \pm 0.004 \pm 0.019$ & +0.05 & $(+1.028 \pm 0.025, -2.418 \pm 0.025)$ & $-95.9 \pm 0.7$ (38) & $-94.4 \pm 0.3$ \\
NGC 6752 & 1 & 21541 & 1016 & $-3.161 \pm 0.002 \pm 0.017$ & $-4.039 \pm 0.002 \pm 0.019$ & +0.34 & $(-3.161 \pm 0.022, -4.027 \pm 0.022)$ & $-26.9 \pm 0.5$ (159) & $-26.0 \pm 0.1$ \\
NGC 6779 & 1 & 2319 & 346 & $-2.013 \pm 0.004 \pm 0.018$ & $+1.614 \pm 0.005 \pm 0.024$ & -0.11 & $(-2.018 \pm 0.025, +1.618 \pm 0.025)$ & $-134.2 \pm 0.8$ (41) & $-137.0 \pm 0.5$ \\
Terzan 7 & 3 & 156 & 46 & $-2.970 \pm 0.015 \pm 0.022$ & $-1.648 \pm 0.015 \pm 0.021$ & +0.14 & $(-3.002 \pm 0.029, -1.651 \pm 0.029)$ & -- & $+159.4 \pm 0.1$ \\
Arp 2 & 4 & 123 & 22 & $-2.319 \pm 0.013 \pm 0.020$ & $-1.488 \pm 0.011 \pm 0.020$ & +0.09 & $(-2.331 \pm 0.031, -1.475 \pm 0.029)$ & -- & $+122.6 \pm 0.3$ \\
NGC 6809 & 1 & 10534 & 811 & $-3.425 \pm 0.003 \pm 0.022$ & $-9.314 \pm 0.003 \pm 0.017$ & -0.12 & $(-3.432 \pm 0.024, -9.311 \pm 0.024)$ & $+174.5 \pm 0.7$ (97) & $+174.7 \pm 0.2$ \\
Terzan 8 & 1 & 371 & 38 & $-2.487 \pm 0.009 \pm 0.019$ & $-1.576 \pm 0.006 \pm 0.018$ & -0.03 & $(-2.496 \pm 0.027, -1.581 \pm 0.026)$ & $+151.8 \pm 4.0$ (4) & $+148.5 \pm 0.2$ \\
Pal 11 & 1 & 288 & 50 & $-1.757 \pm 0.011 \pm 0.023$ & $-4.957 \pm 0.009 \pm 0.026$ & +0.33 & $(-1.766 \pm 0.030, -4.971 \pm 0.028)$ & -- & $-67.6 \pm 0.8$ \\
NGC 6838 & 1 & 3550 & 171 & $-3.412 \pm 0.003 \pm 0.021$ & $-2.659 \pm 0.003 \pm 0.022$ & -0.04 & $(-3.416 \pm 0.025, -2.656 \pm 0.024)$ & $-22.9 \pm 0.4$ (139) & $-22.7 \pm 0.2$ \\
NGC 6864 & 1 & 732 & 113 & $-0.595 \pm 0.007 \pm 0.020$ & $-2.799 \pm 0.006 \pm 0.018$ & +0.02 & $(-0.598 \pm 0.026, -2.810 \pm 0.026)$ & $-185.5 \pm 1.0$ (11) & $-189.1 \pm 1.1$ \\
NGC 6934 & 1 & 1374 & 124 & $-2.657 \pm 0.006 \pm 0.025$ & $-4.686 \pm 0.005 \pm 0.019$ & +0.20 & $(-2.655 \pm 0.026, -4.689 \pm 0.026)$ & $-406.3 \pm 0.9$ (20) & $-406.2 \pm 0.7$ \\
NGC 6981 & 1 & 1148 & 244 & $-1.269 \pm 0.006 \pm 0.018$ & $-3.359 \pm 0.005 \pm 0.018$ & +0.00 & $(-1.274 \pm 0.026, -3.361 \pm 0.026)$ & $-327.5 \pm 0.9$ (12) & $-331.4 \pm 1.5$ \\
NGC 7006 & 1 & 222 & 71 & $-0.126 \pm 0.011 \pm 0.034$ & $-0.644 \pm 0.010 \pm 0.035$ & +0.67 & $(-0.128 \pm 0.027, -0.633 \pm 0.027)$ & -- & $-383.5 \pm 0.7$ \\
NGC 7078 & 1 & 8014 & 669 & $-0.653 \pm 0.004 \pm 0.019$ & $-3.813 \pm 0.003 \pm 0.016$ & +0.14 & $(-0.659 \pm 0.024, -3.803 \pm 0.024)$ & $-106.1 \pm 0.8$ (54) & $-106.8 \pm 0.3$ \\
NGC 7089 & 1 & 4918 & 714 & $+3.445 \pm 0.006 \pm 0.019$ & $-2.169 \pm 0.004 \pm 0.015$ & +0.00 & $(+3.435 \pm 0.025, -2.159 \pm 0.024)$ & $-2.2 \pm 1.1$ (57) & $-3.8 \pm 0.3$ \\
NGC 7099 & 1 & 4430 & 932 & $-0.740 \pm 0.005 \pm 0.025$ & $-7.293 \pm 0.004 \pm 0.019$ & -0.38 & $(-0.737 \pm 0.025, -7.299 \pm 0.024)$ & $-184.6 \pm 0.7$ (30) & $-185.2 \pm 0.2$ \\
Pal 12 & 1 & 301 & 149 & $-3.225 \pm 0.013 \pm 0.021$ & $-3.339 \pm 0.011 \pm 0.018$ & +0.02 & $(-3.220 \pm 0.029, -3.333 \pm 0.028)$ & $+29.0 \pm 1.3$ (4) & $+27.9 \pm 0.3$ \\
Pal 13 & 1 & 58 & 44 & $+1.748 \pm 0.035 \pm 0.027$ & $+0.087 \pm 0.028 \pm 0.029$ & +0.30 & $(+1.748 \pm 0.049, +0.104 \pm 0.047)$ & -- & $+25.3 \pm 0.2$ \\
NGC 7492 & 1 & 280 & 161 & $+0.761 \pm 0.010 \pm 0.019$ & $-2.318 \pm 0.009 \pm 0.020$ & +0.09 & $(+0.756 \pm 0.028, -2.320 \pm 0.028)$ & -- & $-176.7 \pm 0.3$ \\
\bottomrule
\end{tabular}}
\end{table*}

The systemic PMs of the 93 clusters recovered in Sect.~\ref{sec:gcs} are listed in Tables~\ref{tab:gcs_a} and \ref{tab:gcs_b}; the member catalogs and the summary files, which also contain the systemic parallaxes, the fitted profiles and the mean $v_{\rm los}$, are distributed with the code.

\section{Structural parameters of the dwarf galaxies}\label{apx:structure}

The spatial term of the model is an elliptical profile fitted jointly with the membership, so that the run also returns the half-light radius, the ellipticity and the position angle of every system whose profile could be fitted freely: 28 of the 55 recovered dwarfs, 19 from the preliminary fit and nine from the refit within the EM (Sect.~\ref{sec:regions}; for the others the shape of the finder or of the literature was kept fixed). Their uncertainties come from a bootstrap over the members (100 resamplings, refitting the profile each time), which captures the sampling noise of the fit but not the choice of the profile. Figure~\ref{fig:structure} compares them with the photometric structural parameters compiled in the LVDB, which come from deep imaging \citep[mostly][]{Munoz2018}. The five star-forming dwarfs (Phoenix, NGC\,6822, IC\,1613, IC\,10 and Leo\,A) are a case apart. At their distances the RGB of the old population is fainter than $G = 21$, and the members are young supergiants whose distribution need not follow the old light that defines the photometric parameters. NGC\,6822 is the clearest example, with a position angle $85\degree$ from the photometric one; the blue and the red members give the same orientation, so the difference is a property of the population traced, not of the fit. These five systems are kept apart in the figure and in the statistics. For the 23 Milky Way satellites the position angles agree with the literature within $7\degree$ (median) and within $2\sigma$ for 16 systems. The bootstrap errors are $1$--$3\degree$ for the classical dSphs but reach tens of degrees for the systems with a few tens of members or rounder than $\epsilon = 0.25$, where the orientation is intrinsically ill defined. The ellipticities agree within 0.01 (median) and within $2\sigma$ for 17 systems. The half-light radii have a median ratio of 1.21 and a median absolute deviation of 21 percent. Here the bootstrap errors ($0.1$--$2$ arcmin) are far smaller than the differences, which are therefore systematic: the \Gaia{} members are the giants and HB stars of each system and not its full luminosity, the profile is an exponential whereas the photometric values come from several functional forms, and in the crowded centers of the classical dSphs a fraction of the stars fail the quality screening (Fig.~\ref{fig:fornax}), which flattens the observed profile. Draco II, Coma Berenices, Ursa Major I and Leo IV, with 5 to 62 members, come out 2.1 to 3.8 times larger, the expected bias when a few tens of stars with individual membership probabilities constrain a profile from the position alone against a field that is not perfectly uniform; the two first also have position angles $55\degree$ and $45\degree$ from the photometric ones, at $3\sigma$. These parameters are a by-product of the selection and less precise than dedicated photometric studies, but they are obtained from the same astrometrically selected members at no extra cost, and they provide a consistency check of the spatial term for every system in the table.

\begin{table*}[p]
\caption{Structural parameters from the spatial term for the detected dwarf galaxies with a freely fitted profile, compared with the LVDB \citep{Pace2025}: number of members, half-light radius along the major axis, ellipticity and position angle (North through East; ours is given at the representation closest to the literature value). The errors are bootstrap uncertainties over the members. Systems marked with $^\ast$ are star-forming dwarfs whose \Gaia{} members are young stars. A dash marks a quantity not listed in the LVDB.}
\label{tab:structure}
\centering
\scriptsize
\setlength{\tabcolsep}{2.5pt}
\resizebox{\textwidth}{!}{%
\begin{tabular}{l r r r r r r r @{\hspace{14pt}} l r r r r r r r}
\toprule
Galaxy & $N_{\rm mem}$ & $r_h$ & $r_h^{\rm lit}$ & $\epsilon$ & $\epsilon^{\rm lit}$ & PA & PA$^{\rm lit}$ & Galaxy & $N_{\rm mem}$ & $r_h$ & $r_h^{\rm lit}$ & $\epsilon$ & $\epsilon^{\rm lit}$ & PA & PA$^{\rm lit}$ \\
       &  & [$'$] & [$'$] &  &  & [$\degree$] & [$\degree$] &  &  & [$'$] & [$'$] &  &  & [$\degree$] & [$\degree$] \\
\midrule
Draco II & 28 & $11.4 \pm 1.1$ & 3.0 & $0.13 \pm 0.10$ & 0.23 & $21 \pm 16$ & 76 & Ursa Major I & 62 & $22.7 \pm 1.6$ & 8.3 & $0.61 \pm 0.08$ & 0.59 & $58 \pm 6$ & 67 \\
Tucana III & 25 & $6.2 \pm 0.7$ & 5.1 & $0.26 \pm 0.12$ & 0.20 & $75 \pm 26$ & 25 & Carina & 2044 & $14.5 \pm 0.2$ & 10.1 & $0.38 \pm 0.01$ & 0.36 & $64 \pm 2$ & 60 \\
Hydrus I & 86 & $9.9 \pm 1.0$ & 7.4 & $0.11 \pm 0.08$ & 0.21 & $118 \pm 16$ & 97 & Crater II & 200 & $25.8 \pm 1.1$ & 31.2 & $0.03 \pm 0.08$ & -- & $96 \pm 28$ & -- \\
Reticulum II & 50 & $6.1 \pm 0.5$ & 6.3 & $0.63 \pm 0.05$ & 0.60 & $64 \pm 4$ & 68 & Fornax & 19982 & $17.7 \pm 0.1$ & 19.9 & $0.31 \pm 0.01$ & 0.29 & $42 \pm 1$ & 43 \\
Ursa Major II & 38 & $16.7 \pm 2.2$ & 13.8 & $0.50 \pm 0.09$ & 0.56 & $111 \pm 10$ & 104 & Leo IV & 5 & $6.1 \pm 3.9$ & 2.5 & $0.09 \pm 0.20$ & 0.17 & $162 \pm 33$ & 152 \\
Carina II & 37 & $12.0 \pm 2.0$ & 8.7 & $0.35 \pm 0.12$ & 0.34 & $170 \pm 21$ & 170 & Canes Venatici II & 14 & $1.8 \pm 0.3$ & 1.5 & $0.59 \pm 0.19$ & 0.40 & $-3 \pm 30$ & 9 \\
Boötes II & 22 & $4.1 \pm 0.4$ & 3.2 & $0.00 \pm 0.16$ & 0.25 & $180 \pm 50$ & 112 & Canes Venatici I & 142 & $10.5 \pm 0.5$ & 7.1 & $0.29 \pm 0.05$ & 0.44 & $87 \pm 7$ & 80 \\
Coma Berenices & 37 & $11.7 \pm 1.3$ & 5.6 & $0.18 \pm 0.12$ & 0.37 & $78 \pm 16$ & 123 & Leo II & 313 & $2.5 \pm 0.1$ & 2.5 & $0.15 \pm 0.04$ & 0.07 & $45 \pm 12$ & 38 \\
Tucana II & 26 & $19.2 \pm 3.4$ & 12.9 & $0.23 \pm 0.15$ & 0.39 & $74 \pm 24$ & 107 & Leo I & 1191 & $4.0 \pm 0.1$ & 3.6 & $0.39 \pm 0.02$ & 0.30 & $84 \pm 2$ & 78 \\
Sagittarius II & 53 & $1.8 \pm 0.2$ & 1.9 & $0.26 \pm 0.08$ & 0.03 & $89 \pm 28$ & 96 & Phoenix$^{\ast}$ & 207 & $2.0 \pm 0.1$ & 2.4 & $0.33 \pm 0.05$ & 0.30 & $5 \pm 5$ & 8 \\
Draco & 1568 & $9.9 \pm 0.2$ & 9.7 & $0.29 \pm 0.02$ & 0.29 & $91 \pm 2$ & 87 & NGC 6822$^{\ast}$ & 1382 & $6.1 \pm 0.1$ & 11.9 & $0.34 \pm 0.02$ & 0.28 & $152 \pm 2$ & 67 \\
Ursa Minor & 1915 & $21.2 \pm 0.3$ & 18.3 & $0.45 \pm 0.01$ & 0.55 & $50 \pm 1$ & 50 & IC 1613$^{\ast}$ & 584 & $4.2 \pm 0.1$ & 7.6 & $0.26 \pm 0.03$ & 0.20 & $63 \pm 4$ & 90 \\
Sculptor & 6604 & $13.1 \pm 0.2$ & 11.2 & $0.31 \pm 0.01$ & 0.33 & $94 \pm 1$ & 92 & IC 10$^{\ast}$ & 29 & $1.7 \pm 0.6$ & 2.6 & $0.60 \pm 0.17$ & 0.19 & $91 \pm 6$ & -- \\
Sextans & 1307 & $28.0 \pm 0.5$ & 22.8 & $0.23 \pm 0.03$ & 0.27 & $53 \pm 3$ & 52 & Leo A$^{\ast}$ & 46 & $1.1 \pm 0.1$ & 2.3 & $0.39 \pm 0.07$ & 0.42 & $84 \pm 11$ & 116 \\
\bottomrule
\end{tabular}}
\end{table*}

\section{Installation and use}\label{apx:usage}

\getgaia{} 2.0 requires Python 3.10 or later and is installed with \texttt{pip install -e .[dr3]} from the repository; the \texttt{[dr3]} extra installs the \texttt{gaiadr3-zeropoint} package. A complete run on a named object is

\begin{verbatim}
getgaia --name "Fornax dSph" --quasars
\end{verbatim}

\noindent which resolves the object with Simbad, downloads the data (as synchronous queries over \texttt{source\_id} intervals; the page size and the number of concurrent queries are options, and a page that fails is resubmitted as an asynchronous job), fits the model and writes, in the directory \texttt{Fornax\_dSph/}, the full catalog with the probabilities, the list of members, a JSON summary with the systemic motion and its covariance matrices, the structural parameters and the quasar analysis, two diagnostic figures and a log. The most relevant options are \texttt{--release} (\texttt{dr3} or \texttt{dr4}), \texttt{--search-radius} and \texttt{--field-radius} (degrees), \texttt{--max-gmag}, \texttt{--threshold} (membership probability), \texttt{--distance} (kpc) and \texttt{--sigma-v-max} (cap on the intrinsic dispersion), \texttt{--no-sky}, \texttt{--no-cmd} and \texttt{--no-parallax} (disable terms of the model), \texttt{--field-components}, \texttt{--no-systematics} or \texttt{--pm-systematic}, \texttt{--blind} (position only; Sect.~\ref{sec:detection}), \texttt{--pm-fallback} and \texttt{--no-ask} (initial PM when Simbad lists none), \texttt{--annulus-contrast} and \texttt{--annulus-inner} (object/field boundary), \texttt{--half-light-radius}, \texttt{--ellipticity} and \texttt{--position-angle} (literature structural parameters), \texttt{--no-iterate-regions} (single pass of the regions), \texttt{--quasar-radius}, \texttt{--quasar-correct} and \texttt{--quasar-cone-only}, \texttt{--rv-catalog} with \texttt{--rv-columns}, \texttt{--rv-match-radius} and \texttt{--rv-epoch} (external line-of-sight velocities), \texttt{--no-rv}, \texttt{--no-quality-cut} and the individual quality thresholds, and \texttt{--interactive}, which opens a slider on the probability threshold and a lasso on the CMD to refine the selection by hand. The same pipeline is available from Python through \texttt{getgaia.run(PipelineConfig(...))}, and the membership model alone through \texttt{getgaia.membership.MembershipModel}, which accepts any table with the required columns. A script to process a list of objects and compile the results is provided in the \texttt{examples} directory.

\end{document}